\documentclass[conference]{IEEEtran}

\usepackage{mathptmx} 
\usepackage[utf8]{inputenc}
\usepackage[T1]{fontenc}
\usepackage{microtype}
\usepackage{enumitem}
\setlist[itemize]{leftmargin=*, topsep=0pt}
\usepackage{pifont}
\usepackage[dvipsnames,table,xcdraw]{xcolor}
\usepackage{tabularx}

\newif\ifremark
\long\def\remark#1{
\ifremark%
        \begingroup%
        \dimen0=\columnwidth
        \advance\dimen0 by -1in%
        \setbox0=\hbox{\parbox[b]{\dimen0}{\protect\em #1}}
        \dimen1=\ht0\advance\dimen1 by 2pt%
        \dimen2=\dp0\advance\dimen2 by 2pt%
        \vskip 0.25pt%
        \hbox to \columnwidth{%
                \vrule height\dimen1 width 3pt depth\dimen2%
                \hss\copy0\hss%
                \vrule height\dimen1 width 3pt depth\dimen2%
        }%
        \endgroup%
\fi}

\usepackage{listings}
\usepackage{parcolumns}

\definecolor{dkgreen}{rgb}{0,0.6,0}
\definecolor{gray}{rgb}{0.5,0.5,0.5}
\definecolor{mauve}{rgb}{0.58,0,0.82}

\remarktrue

\newcommand{\pcm}{PCM-Only\xspace}
\newcommand{\wrgc}{Kingsguard\xspace}

\newcommand{\basic}{\wrgc-nursery\xspace}

\newcommand{\writers}{\wrgc-writers\xspace}

\newcommand{\kgn}{KG-N\xspace}
\newcommand{\kgw}{KG-W\xspace}
\newcommand{\kgb}{KG-B\xspace}
\newcommand{\kgp}{KG-B\xspace}

\usepackage{amsmath}
\newcommand{\ignore}[1]{}

\usepackage{graphicx}
\usepackage{subfig}

\usepackage[ruled,norelsize]{algorithm2e}

\usepackage{hyperref}
\usepackage{url}

\usepackage{fancyhdr}
\usepackage[normalem]{ulem}
\usepackage{array}
\usepackage{bbding}

\usepackage{booktabs}

\usepackage{array}
\usepackage{arydshln}
\usepackage[T1]{fontenc}
\usepackage[labelfont=bf,tableposition=top]{caption}

\DeclareCaptionLabelFormat{andtable}{#1~#2  \&  \tablename~\thetable}

\usepackage[normalem]{ulem}

\begin{document}

\title{Emulating Hybrid Memory on NUMA Hardware}

\author{\IEEEauthorblockN{Shoaib Akram}
\IEEEauthorblockA{
Ghent University\\
}
\and
\IEEEauthorblockN{Jennifer B. Sartor}
\IEEEauthorblockA{
Ghent University \\
}
\and
\IEEEauthorblockN{Kathryn S. McKinley}
\IEEEauthorblockA{
Google\\
}
\and
\IEEEauthorblockN{Lieven Eeckhout}
\IEEEauthorblockA{
Ghent University\\
}
}

\maketitle
\thispagestyle{plain}
\pagestyle{plain}

\begin{abstract}
  Non-volatile memory (NVM) has the potential to disrupt the boundary
  between memory and storage, including the abstractions that manage
  this boundary.  Researchers comparing the speed, durability, and
  abstractions of hybrid systems with DRAM, NVM, and disk to
  traditional systems typically use simulation, which makes it easy to
  evaluate different hardware technologies and
  parameters. Unfortunately, simulation is extremely slow, limiting
  the number of applications and dataset sizes in the evaluation.
  Simulation typically precludes realistic multiprogram workloads and
  considering runtime and operating system design alternatives.

  Good methodology embraces a variety of techniques for validation,
  expanding the experimental scope, and uncovering new insights.  This paper
  introduces an emulation platform for hybrid memory that uses
  commodity NUMA servers. Emulation complements simulation well,
  offering speed and accuracy for realistic workloads, and richer
  software experimentation. We use a thread-local socket to emulate
  DRAM and the remote socket to emulate NVM. We use standard C library
  routines to allocate heap memory in the DRAM or NVM socket for use
  with explicit memory management or garbage collection. We evaluate
  the emulator using various configurations of write-rationing garbage
  collectors that improve NVM lifetimes by limiting writes to NVM, and use 
  15 applications from three benchmark suites with various datasets
  and workload configurations. We show emulation enhances simulation
  results.  The two systems confirm most trends, such as NVM write and
  read rates of different software configurations, increasing our
  confidence for predicting future system effects. In a few cases,
  simulation and emulation differ, offering opportunities for
  revealing methodology bugs or new insights. Emulation adds novel
  insights, such as the non-linear effects of multi-program workloads
  on write rates.  We make our software infrastructure publicly
  available to advance the evaluation of novel memory management
  schemes on hybrid memories.

\end{abstract}

\vspace*{-0.5ex}
\section{Introduction}
\label{intro}

Systems researchers and architects have long pursued bridging the speed gap
between processor, memory, and storage. Despite many efforts, the increase in
processor performance has consistently outpaced memory and storage speeds.
Recent advances in memory technologies have the potential to disrupt this speed
gap. 

On the storage side, emerging non-volatile memory (NVM) technologies  with
speed closer to DRAM and persistence similar to disk promise to narrow the
speed gap between processors and storage. \ignore{Recent work integrates NVM in
the storage hierarchy by engineering new filesystem abstractions, storage
stacks, programming models, wear-out mitigation schemes, and prototyping
platforms~\cite{Dulloor:2014:SSP,Condit:2009:BIT,Coburn:2011:NMP,Zhang:2015:MRH,Xu:2016:NLF,
Xu:2017:NFN,Volos:2011:MLP}.} Recent work engineers new filesystem
abstractions, storage stacks, programming models, wear-out mitigation schemes,
and prototyping platforms to integrate NVM in the
storage hierarchy~\cite{Dulloor:2014:SSP,Condit:2009:BIT,Coburn:2011:NMP,Zhang:2015:MRH,Xu:2016:NLF,Xu:2017:NFN,Volos:2011:MLP}.

On the main memory side, NVM promises abundant memory.  DRAM is facing scaling
limitations~\cite{Lim:2009,Mutlu:204:DRAM:Scaling}, and recent work combines
DRAM and NVM to form hybrid main memories~\cite{lee-pcm-isca,moin-pcm-isca}.
DRAM is fast and durable whereas NVM is dense and has low energy. Hardware
mitigates NVM wear-out in both its storage and memory roles using wear-leveling
and other approaches~\cite{moin-pcm-isca,lee-pcm-isca,Moin-Start-Gap,
Moin-Secure-PCM,Seznec-Secure-PCM}, while the OS keeps frequently accessed data
in
DRAM~\cite{ricardo-os,sally-semantics-hybrid,Zhang:2009,Oskin:2015:SAD,Mutlu:2017:UBH,Liu:2016:Memos}.
Recent work also explores managed runtimes to mitigate
wear-out~\cite{PLDI:2018:Shoaib,PASS:2018:Shoaib}, tolerate
faults~\cite{Gao:2013:UMR}, and keep frequently read objects in
DRAM~\cite{Wang:2016}. Collectively, prior research illustrates 
the substantial opportunities to exploit NVM across all layers including
software and language runtimes.

In this paper, we expand on the methodologies for evaluating NVM and hybrid
memories.  The dominant evaluation methodology in prior work is simulation; see
for example~\cite{moin-pcm-isca,lee-pcm-isca,Moin-Start-Gap,
Moin-Secure-PCM,Seznec-Secure-PCM,ricardo-os,sally-semantics-hybrid,Zhang:2009,Mutlu:2017:UBH}.
A few researchers have complemented simulation with architecture-independent
measurements~\cite{Sam:TC,PLDI:2018:Shoaib,PASS:2018:Shoaib}, but these
measurements have limited value because they miss important effects such as
CPU caching. This paper shows emulation confirms the results of simulation 
and architecture-independent analysis and enables researchers to explore
richer software configurations.

\begin{table}[b]
	\centering
	\vspace*{-1.2em}
	\includegraphics[width=8cm]{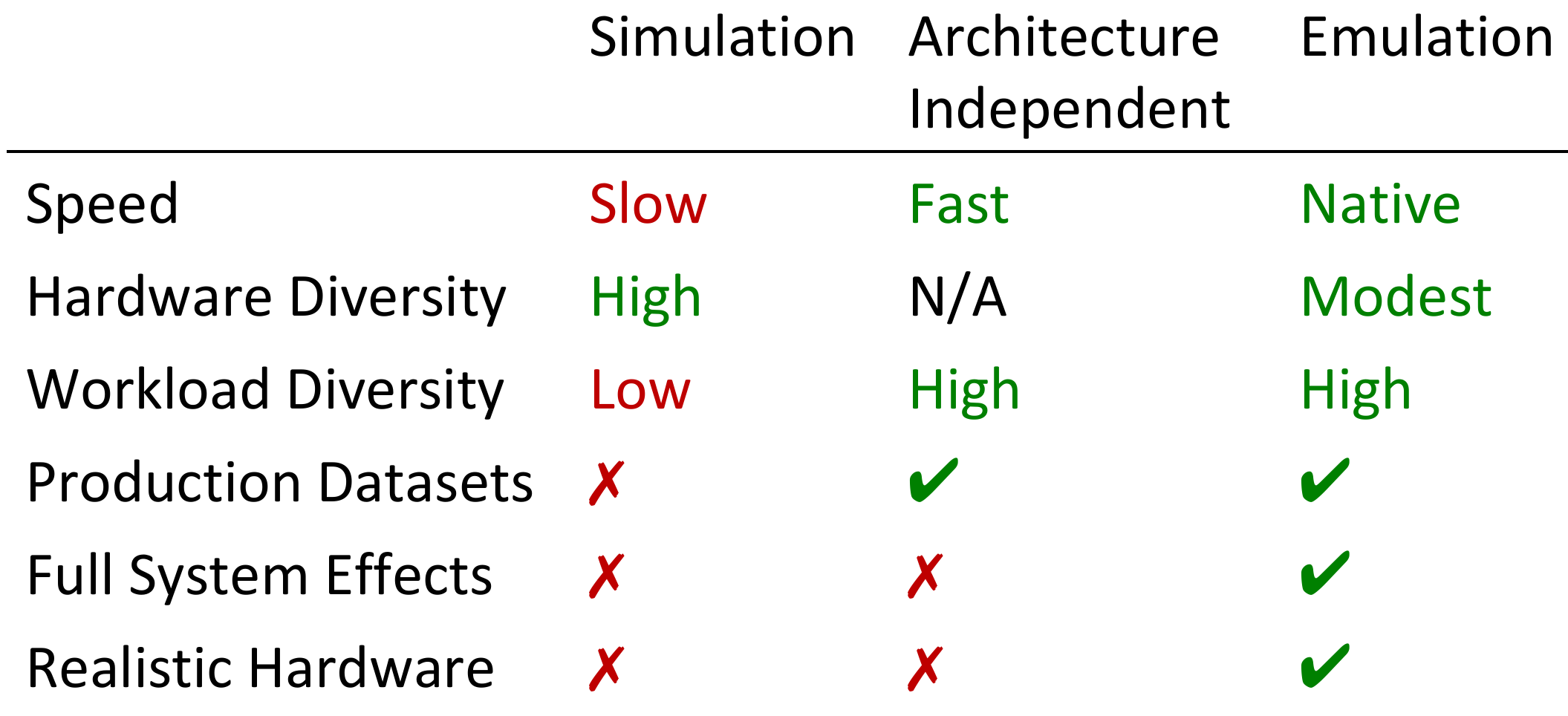}
	\vspace{0.5mm}
	\caption{ Comparing the strengths and weaknesses of evaluation
methodologies for hybrid memories. \textit{Emulation enables native exploration
of diverse workloads and datasets on realistic hardware.}
        }
        \label{tab:comp-methods}
\end{table}

The advantage of simulation is that it eases modeling new hardware features,
revealing how sensitive results are to architecture. Its major limitation is
that it is many orders of magnitude slower than running programs on real
hardware. Because time and resources are finite, it thus reduces the scope and
variety of architecture optimizations, application domains, implementation
languages, and datasets one can explore. Popular simulators also
trade off accuracy to speed up
simulation~\cite{carlson2014aeohmcm,Sanchez:2013:ZFA}.  Furthermore, frequent
hardware changes, microarchitecture complexity, and hardware's proprietary
nature make it difficult to faithfully model real hardware.

Other research evaluations are increasingly embracing emulation. For instance,
emulating cutting-edge hardware on commodity machines to model: asymmetric
multicores using frequency scaling~\cite{yinyang,Haque:2017}, die-stacked and
hybrid memory using DRAM~\cite{Oskin:2015:SAD,Wang:2016,Dulloor:2014:SSP}, and
wearable memory using fault injection software~\cite{Gao:2013:UMR}.
Recent work using emulation for exploring hybrid memory is either limited to
native languages~\cite{Oskin:2015:SAD,Dulloor:2014:SSP}, or is limited to
simplistic heap organizations in the case of managed languages~\cite{Wang:2016}
(See also Section~\ref{sec:related}). Table~\ref{tab:comp-methods} compares the
methodologies for evaluating hybrid memories, showing all can lead to insight
and that emulation has distinct advantages in speed and software
configuration.

We present the design, implementation, and evaluation of an emulation platform
that uses widely available commodity NUMA hardware to model hybrid
DRAM-NVM systems. \ignore{We show how emulation results complement simulation
results.} We use the local socket to emulate DRAM and the remote socket to
emulate NVM. All threads execute on the DRAM socket.  Our heap splits virtual
memory into DRAM and NVM virtual memory, which we manage using two free lists,
one for each NUMA node by explicitly specifying where to allocate memory in the
C standard library. We expose this hybrid memory to the garbage collector,
which directs the OS where in memory (which NUMA node) to map heap regions.
Contrary to most prior work, our platform handles both manual memory management
routines from the standard C library and memory management using an automatic
memory manager (garbage collector). We redesign the memory manager in the
popular Jikes research virtual machine (RVM) to add support for hybrid
memories. Our software infrastructure is publicly available at
\url{<link-anonymized-for-blind-reviewing>}.

We evaluate this emulation platform on recently proposed write-rationing
garbage collectors for hybrid memories~\cite{PLDI:2018:Shoaib}. Write-rationing
collectors keep highly mutated objects in DRAM in hybrid DRAM-NVM systems to
target longer NVM lifetime. We use 15 applications from three benchmark suites:
DaCapo, Pjbb, and GraphChi; two input datasets; seven garbage collector
configurations; and workloads consisting of one, two, and four application
instances executing simultaneously.  We find emulation results are very similar
to simulation results and platform-independent measurements in most cases, but
we can generate a lot more of them in the same amount of time and explore much
richer software configurations and workloads.  The emulator reveals trends not
identified previously by simulation and platform-independent measurements. We
summarize our key findings below.

\begin{itemize}

\item Simulation, emulation, and architecture-independent analysis reveal
similar trends in write rate reductions and other characteristics of garbage
collectors designed for hybrid memories, increasing our confidence in all the
evaluation methodologies.

\item Managed workloads use a lot of C/C++ code. Garbage collection strategies
for hybrid memories should protect against both writes to the managed heap and
writes to memory allocated using explicit C and C++ allocators.


\item Executing multiple applications simultaneously super-linearly increases
NVM write rates due to LLC interference, a configuration that is not practical
to explore in simulation. A major portion of the additional writes to memory
are due to nursery writes. \wrgc collectors isolate these writes on DRAM and
thus are especially effective in multiprogrammed environments.

\item Modern graph processing workloads use larger heaps and their write rates
are also higher than widely used Java benchmarks. Future work should include such
benchmarks when evaluating hybrid memories.

\item Addressing large objects' behaviors are essential to memory
managers for hybrid memories.  Graph applications can see huge reductions in
write rates when using \wrgc collectors, because they have a
lot of large objects that benefit from targeted optimizations. 

\item Changing a benchmark's allocation behavior or input changes write
rates. Future work should eliminate useless allocations and use a variety of
inputs for evaluating hybrid memories.


\item LLC size impacts write rates. Future work should use suitable workloads
with emulation on modern servers with large LLCs, or report evaluation
for a range of LLC sizes using simulation.

\item Graph applications wear PCM out faster than traditional Java benchmarks.
Multiprogramming workloads can also wear PCM out in less than 5 years. Write
limiting with \wrgc collectors brings PCM lifetimes to practical levels. 

\end{itemize}

\vspace*{-1.3ex}
\section{Background}
\label{sec:background}

This section briefly discusses characteristics of NVM hardware and the role of
DRAM in hybrid DRAM-NVM systems. We then discuss write-rationing garbage
collection~\cite{PLDI:2018:Shoaib} that protect NVM from writes and prolongs
memory lifetime. We will evaluate write-rationing garbage collectors in
Section~\ref{sec:results} using our emulation platform.

\vspace*{-0.8ex}
\subsection{NVM Drawbacks and Hybrid Memory}

A promising NVM technology currently in production is phase change
memory (PCM)~\cite{xpoint}. PCM cells store information as the change in
resistance of a chalcogenide material~\cite{PCM-Vacum-Science}.  During a write
operation, electric current heats up PCM cells to high temperatures and the
cells cool down into an amorphous or a crystalline state that have different
resistances. The read operation simply detects the resistance of the cell.  PCM
cells wear out after 1 to 100 million writes because each write changes their physical
structure~\cite{lee-pcm-isca,moin-pcm-isca,PCM-Vacum-Science}.  Writes are also
an order of magnitude slower and consume more energy than in DRAM. Reading the PCM
array is up to 4$\times$ slower than DRAM~\cite{lee-pcm-isca}.  

Hybrid memories combine DRAM and PCM to mitigate PCM wear-out and tolerate its
higher latency. Frequently accessed data is kept in DRAM which results in better
performance and longer lifetimes compared to a \pcm system. The large PCM
capacity reduces disk accesses which compensates for its slow speed.

\vspace*{-0.7ex}
\subsection{Garbage Collection}

\paragraph{Generational Garbage Collection}
Managed languages such as Java, C\#, Python, and JavaScript use garbage
collection to accelerate development and reduce memory errors.
High-performance garbage collectors today exploit the generational hypothesis
that most objects die young~\cite{Ungar:1984:GSN}.  With generational
collectors, applications (mutators) allocate new objects contiguously into a
nursery. When allocation exhausts the nursery, a minor collection first
identifies live roots that point into the nursery, e.g., from global variables,
the stack, registers, and the mature space. It then identifies reachable
objects by tracing references from these roots. It copies reachable objects to
a mature space and reclaims all nursery memory for subsequent fresh allocation.
When the mature space is full, a full-heap (mature) collection collects the
entire heap. Recent work exploits garbage collectors to manage hybrid
memories~\cite{PLDI:2018:Shoaib,Gao:2013:UMR} and to improve PCM lifetimes.

\paragraph{Write-Rationing Garbage Collection}

Write-rationing collectors keep frequently written objects in DRAM in 
hybrid memories to improve PCM lifetime~\cite{PLDI:2018:Shoaib}. They come
in two main variants: The \emph{\basic} (\kgn) collector allocates nursery
objects in DRAM and promotes all nursery survivors to PCM. The nursery is highly
mutated and \kgn reduces write rates significantly compared to \pcm which leads
to a longer PCM lifetime.  \emph{\writers} (\kgw) monitors nursery survivors in
a DRAM observer space.  Observer space collections copy objects with zero
writes to a PCM mature space, and copy written objects to a DRAM mature space.
\kgw incurs a moderate performance overhead over \kgn due to monitoring and
extra copying of some nursery survivors but further improves PCM lifetime over
\kgn.

\kgw includes two additional optimizations to protect PCM from writes.
Traditional garbage collectors allocate large objects directly in a non-moving
mature space to avoid copying them from the nursery to the mature space.
Large Object Optimization (LOO) in \kgw allocates some large objects, chosen using a
heuristic, in the nursery giving them time to die. Like standard collectors, the
mutator allocates the remaining large objects directly in a PCM mature space.
The collector copies highly written large objects from PCM to DRAM during a
mature collection.  Garbage collectors also write to object metadata to mark
them live.  Marking live objects generates writes to PCM during a mature
collection. MetaData Optimization (MDO) places PCM object metadata in DRAM to
eliminate garbage collector writes to object metadata.

\wrgc collectors build on the best-performing collector in Jikes RVM:
generational Immix (GenImmix)~\cite{immix-blackburn}. GenImmix uses a copying
nursery and a mark-region mature space.

\vspace*{-0.6em}
\section{Design and Implementation}
\label{design}

This section describes the design and implementation of our emulator
for hybrid memory systems. It includes a hybrid-memory-aware memory manager
built on top of the standard NUMA hardware platforms widely available today.


\vspace*{-1ex}
\begin{figure}[bt]
 	\centering
	\includegraphics[width=7.8cm]{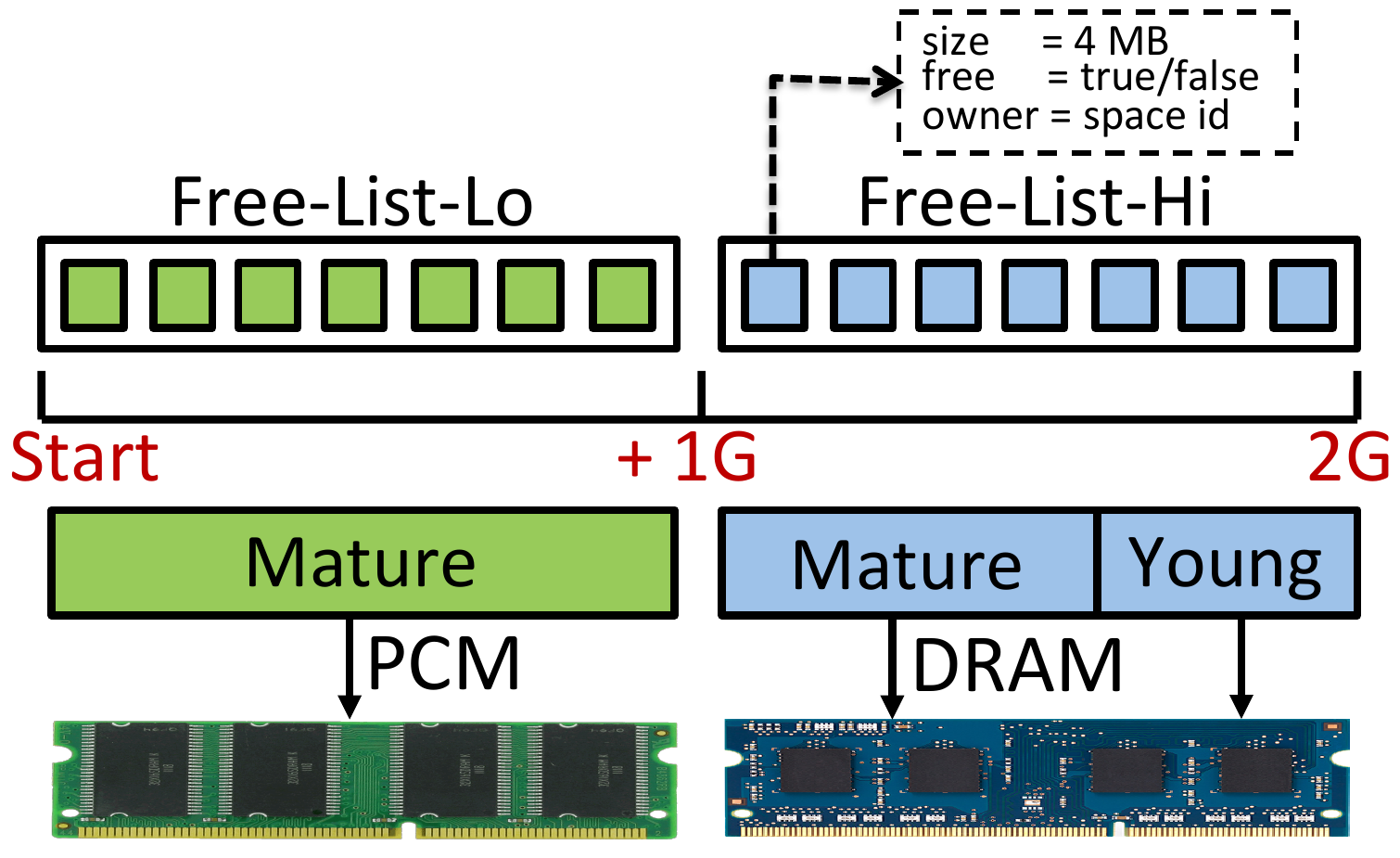}
	\caption{The organization of our heap in hybrid memory.
        \textit{Memory composition is exposed to the language runtime. 
        Two free lists keep track of available virtual pages in DRAM and PCM.}
        }
	\vspace*{-1.5em}
	\label{fig:heap}
\end{figure}

\vspace*{-0.2em}
\subsection{Heap layout and management}

We allocate memory using the Linux OS calls for specifying a memory allocation
on a local or remote memory socket on a NUMA machine. We use the local socket
as the DRAM socket and the remote socket as the PCM socket.  We use a NUMA
specific version of the C memory allocator to call these routines. We modify
the Java Virtual Machine to call the C routines for DRAM and PCM allocation.
Figure~\ref{fig:heap} shows the high-level layout of our heap in hybrid memory.

We use Jikes RVM, but our approach generalizes to other JVMs. Jikes RVM is a
32-bit virtual machine, and each program has 4\:GB of virtual memory. The Linux
OS and system libraries use the low virtual memory for its own purposes. We use
the upper 2\:GB heap for the Java heap. This memory is sufficient for our
applications, although it is possible to use more than 2\:GB.  We partition the
heap into two parts. Each 1\:GB portion is logically divided into 4\:MB chunks
and managed independently by a free-list data structure. Figure~\ref{fig:heap}
shows {\sf Free-List-Hi} and {\sf Free-List-Lo} that keep track of free DRAM
and PCM memory respectively.  Each entry in the free-list contains
meta-information about the chunk: (1) size of the chunk, (2) status of the
chunk (free or in use), and (3) the current owner of the chunk. The lower 1\:GB
portion in virtual memory maps to PCM, and the upper portion maps to DRAM. 

Jikes RVM includes a memory management tool kit (MMTk) to manage the Java
heap.  Standard MMTk configurations flexibly manage portions of the heap using
different allocation and collection mechanisms. Each such portion is called a
space in MMTk terminology. For example, the nursery is a contiguous space and uses
a bump pointer allocator, and its memory is reclaimed using copying collection.
Each space in our implementation reserves virtual memory by requesting the
allocator associated with {\sf Free-List-Lo} or {\sf Free-List-High}. The
allocator finds a free chunk and returns the address to the requesting space.
The space then makes sure the chunk is mapped in physical memory. In our
approach, once a chunk is mapped in physical memory, we do not remove
its mapping in the OS page tables even if the chunk is no longer in use by the
requesting space. The chunk is recycled by the allocator when another space
requests a free chunk. We modify the chunk allocator to map memory on DRAM or PCM.

Alternative approaches are possible although their efficiency might be low.
For instance, a monolithic Java heap with a single free-list would require
unmapping free chunks from the physical memory. Because otherwise, a DRAM space
could end up using a logical chunk that is physically mapped in PCM.  The
flexibility of leaving the free chunks mapped in physical memory is a result of
our design with two free lists. 

Spaces such as the nursery have their address ranges reserved at boot-time.
On the other hand, mature spaces use a request mechanism to acquire chunks. These spaces share the
pool of available chunks with other spaces. Both types of spaces can be placed
in either DRAM or PCM. Each space specifies  DRAM or NVM as a flag in its
constructor.

Similar to the baseline design, we place the young generation (nursery) at one
side of the virtual memory.  This configuration enables the standard fast
boundary write-barrier for generational collection. Other contiguous spaces
such as the observer space in \kgw are placed next to the nursery. 

MMTk uses {\sf mmap()} for reserving virtual memory if none is available as
indicated by the free lists. To bind a virtual memory range to a particular
socket, we call {\sf mbind()} with the socket number after each call to {\sf
mmap()}.

\vspace*{-0.4em}
\subsection{Emulation on NUMA Hardware}

\paragraph*{\textbf{Hardware requirements}}

\begin{figure}[b]
	\centering
	\vspace*{-1.5em}  
	\includegraphics[width=8cm]{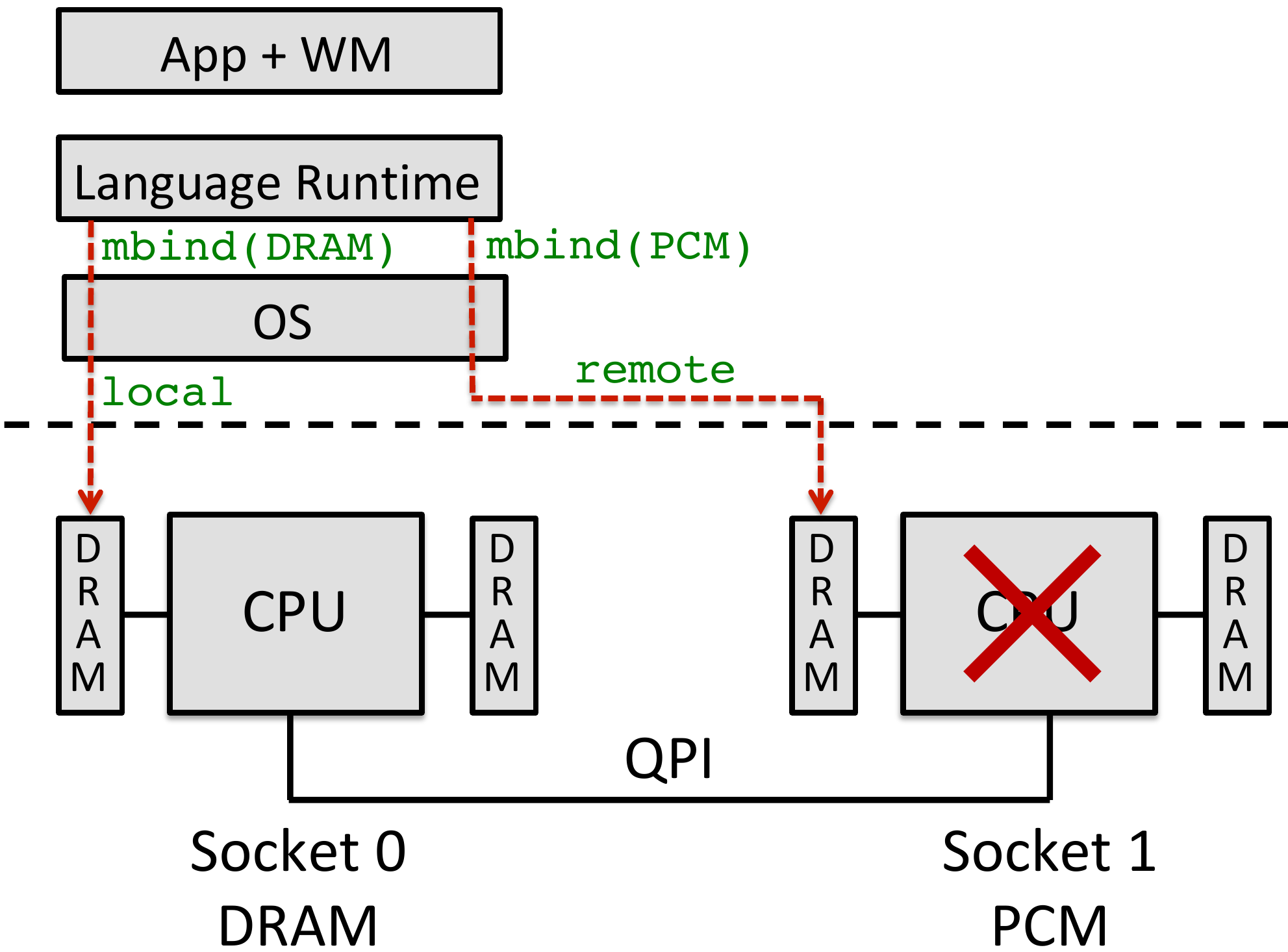}
	\caption{
		Our platform for hybrid memory emulation. \textit{The application
		and write rate monitor (WM) runs on socket 0. The memory on socket 0 is DRAM
		and socket 1 is PCM.}
          }
        \label{fig:platform}
\end{figure}

Our hardware requirement is a commodity NUMA platform with two sockets. We
require both sockets be populated with DRAM chips. Threads run on one socket,
referred to as the local DRAM socket. No threads execute on the other remote
PCM socket. Figure~\ref{fig:platform} shows an example NUMA hardware platform.
Allocation on Socket 0 (S0) is local to the threads and we use it to allocate
DRAM memory.  Memory accesses on Socket 1 (S1) are remote and emulates PCM.

\paragraph*{\textbf{Space to Socket Mapping}}

\begin{table}[bt]
 	\centering
	\includegraphics[width=7.8cm]{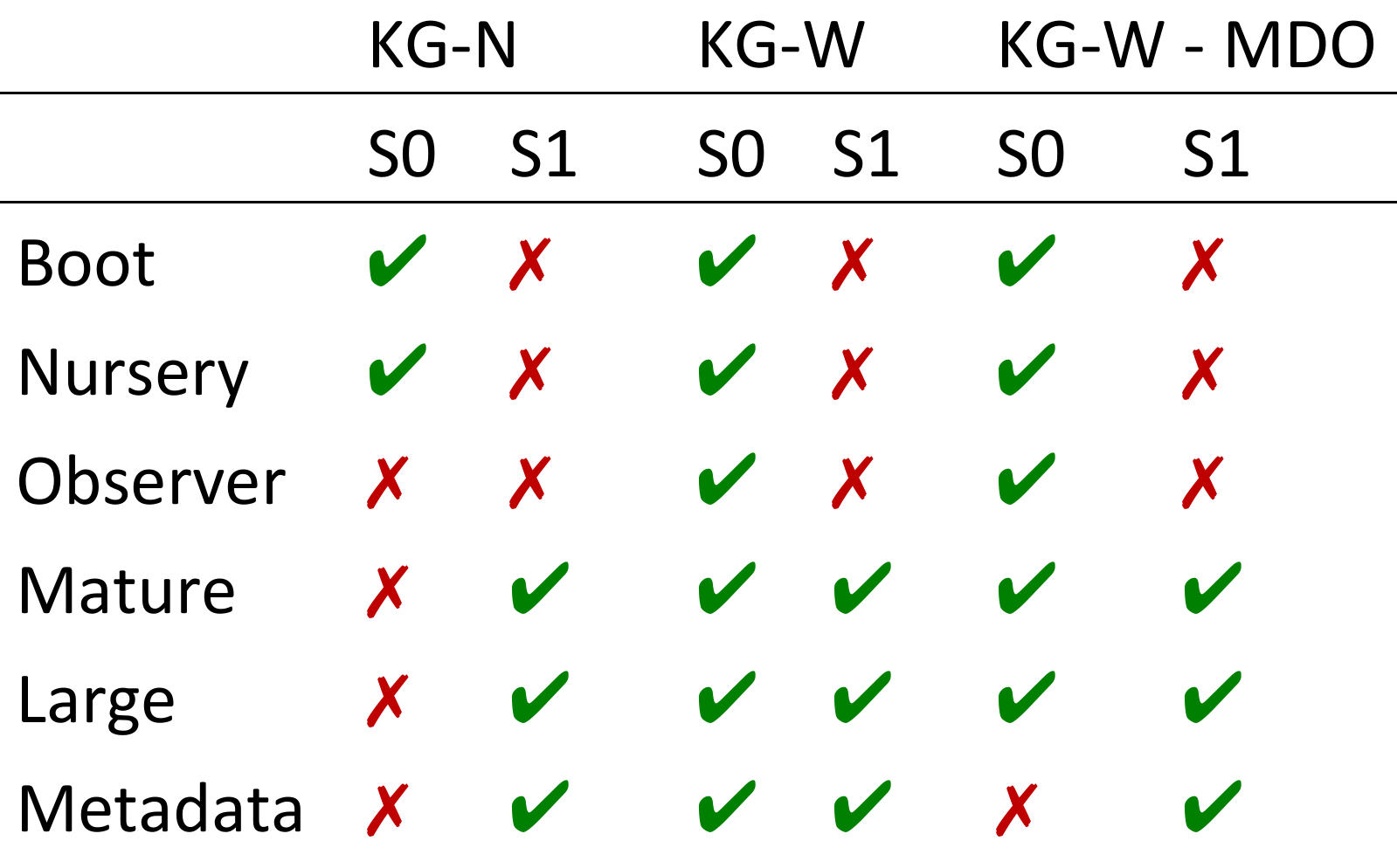}
	\caption{
		 Spaces in \wrgc collectors and their mapping to
socket 0 (S0) or socket 1 (S1). S0 is DRAM and S1 is PCM. \kgn does
not use an observer space. \kgw uses a mature, large, and metadata
space in both DRAM and PCM.  \textit{Our virtual heap layout enables
a range of collector configurations for hybrid memory.}
        }
	\vspace*{-2.5em}  
	\label{fig:heap-maps}
\end{table}

Table~\ref{fig:heap-maps} shows the space to socket mapping in three of the
collectors we evaluate in this work on our emulation platform. 
\kgw and its variants use extra spaces in DRAM that are mapped to
socket 0 (S0). The observer space in \kgw is placed in DRAM and used to monitor
object writes. \kgw has a mature, large, and metadata space in both DRAM (S0)
and PCM (S1). \kgw--MDO does not include the metadata optimization (see
Section~\ref{sec:background}). Therefore, it does not use an extra metadata space
in DRAM.

The boot space contains the boot image runner that boots Jikes RVM and loads
its image files. Except for \pcm, we always place the boot image in DRAM
because we observe a large number of writes to it.

\paragraph*{\textbf{Thread to Socket mapping}}

When a particular thread uses the C or C++ library to allocate memory, the OS
places that memory on the socket where the thread is executing. Thus we have to
control to which socket each thread is mapped.  Our JVM calls down to these  C
and C++ libraries for allocation.  For the \wrgc configurations, we always bind
threads, including application and JVM service threads, to socket 0 (see
Figure~\ref{fig:platform}). \ignore{ Because the application executes on socket 0,
memory requests to socket 0 are local (DRAM), and memory accesses to socket 1
are remote (PCM).} When emulating a \pcm system, we bind threads to socket 1 for
accurately reporting write rates. We do not pin threads to specific cores and
use the default OS scheduler.

This work focuses on PCM lifetimes. PCM lifetime in years depends directly on
its write rate.  We measure write rates on our emulation platform using a write
rate monitor (WM in Figure~\ref{fig:platform}) that also runs on socket 0.
Threads are not pinned to specific cores and we use the default OS scheduler.
We experimentally find out that scheduling WM on socket 0 leads to more
deterministic write rate measurements. When scheduled on socket 1 and all
allocation isolated to socket 0, we continue to observe memory traffic on
socket 1.  

\section{Experimental Methodology}
\label{sec:methodology}

\paragraph*{\textbf{Java Virtual Machine}}

We use Jikes RVM 3.1.2 because it uses software
practices that favor ease of modification, while still delivering good
performance~\cite{AAB+:05, alpern_et_al_2000, Oil:04,Frampton:2009}.  As a
comparison point, it took Hotspot~\cite{url:hs,hotspot} 10 years from the
publication of the G1 collector~\cite{Detlefs:2004} to its release. Jikes RVM
is a Java-in-Java VM with both a baseline and a just-in-time optimizing
compiler, but lacks an interpreter. Jikes RVM has a wide variety of garbage
collectors~\cite{mmtk,immix-blackburn,SBYM:13}. Its memory management tool kit
(MMTk)~\cite{mmtk} makes it easy to compose new collectors by combining
existing modules and changing the calls to the C and OS allocators. Jikes RVM also offers easy-to-modify write
barriers~\cite{YBFH:12} which makes it easy to implement a range of 
heap organizations.

\paragraph*{\textbf{Evaluation Metrics}}

We use two metrics to evaluate write-rationing garbage collectors: write
rate and execution time. PCM lifetime is directly proportional to its
write rate~\cite{moin-pcm-isca,Qureshi:2011:PLH,Qureshi:2009:ELS}. We report
execution time both for single application and multiprogrammed workloads.  Our
multiprogrammed workloads consist of multiple instances of the same
application.  On a properly provisioned platform, all instances should finish
execution at the same time. However, due to shared resources, there is
variation in the execution time of individual instances. We find
the variation on our platform to be low.

\paragraph*{\textbf{Measurement Methodology}}

We use best practices from prior work for evaluating Java applications on our
emulation platform~\cite{ha_et_al_2008,huang_et_al_2004}. To eliminate
non-determinism due to the optimizing compiler, we use replay compilation as used
in prior work. Replay compilation requires two iterations of a Java
application in a single experiment. During the first iteration, the VM
compiles each method to a pre-determined optimization level recorded in a prior
profiling run. The second measured iteration does not recompile methods leading
to steady-state behavior. We perform each experiment four times and report the
arithmetic mean. 


We use Intel's Performance Counter Monitor framework for measuring write rates.
We use the {\sf pcm-memory} utility in the framework for measuring write rates.
We make modest modifications to support multiprogrammed
workloads and to make it compatible for use with replay compilation. In a
multiprogrammed workload, all applications synchronize at a barrier and start
the second iteration at the same time.  

\paragraph*{\textbf{Java Applications}}

We use 15 Java applications from three diverse sources: 11
DaCapo~\cite{blackburn_et_al_2006}, pseudojbb2005 (Pjbb)~\cite{pjbb}, and 3
applications from the GraphChi framework for processing graphs~\cite{graphchi}.
The GraphChi applications we use are: (1) {\sf page rank (PR)}, (2) {\sf
connected components (CC)}, and (3) {\sf ALS matrix factorization (ALS)}.
Compared to recent work~\cite{PLDI:2018:Shoaib}, we drop {\sf jython} as it
does not execute stably with our Jikes RVM configuration.  To improve benchmark
diversity, we use updated versions of {\sf lusearch} and {\sf pmd} in addition
to their original versions. {\sf lu.Fix} eliminates useless
allocation~\cite{YBF+:11}, and {\sf pmd.S} eliminates a scalability bottleneck
in the original version due to a large input file~\cite{DuBoisOopsla13}. Similar
to recent prior work, we run the multithreaded DaCapo applications, Pjbb, and
GraphChi applications with four application threads. 

Unless otherwise stated, we use the default datasets for DaCapo and Pjbb. Our
default dataset for GraphChi is as follows: for {\sf PR} and {\sf CC}, we
process 1\:M edges using the LiveJournal online social network~\cite{url:snap},
and for {\sf ALS}, we process 1\:M ratings from the training set of the Netflix
Challenge. The DaCapo suite comes packaged with large datasets for a subset of
the benchmarks. Our large dataset for GraphChi consists of 10\:M edges and
10\:M ratings.

Even though we do not include C and C++ benchmarks in this work, many of our
Java benchmarks exhibit the common behavior of mixing some C/C++ with Java
because Java standard features, such as IO, use C implementations.  For
example,  the DaCapo benchmarks execute a lot of C code~\cite{blink}.

\paragraph*{\textbf{Workload Formation}}

Multiprogrammed workloads reflect real-world server workloads because: (1) A
single application does not always scale with more cores, and (2)
multiprogramming helps amortize server real-estate and cost. Our
multiprogrammed workloads consist of two and four instances of the same
application. We do not restart applications after they finish execution.  To
avoid non-determinism due to sharing in the OS caches in multiprogrammed
workloads, we use independent copies of the same dataset for the different
instances.

\paragraph*{\textbf{Garbage Collectors and Configurations}}

We explore seven write-rationing garbage collectors.\ignore{ shown in
Table~\ref{tab:cfgs}.} 

Our collector configurations include \kgn, and a variant called
\kgb, that uses a bigger nursery than \kgn. \kgb and its variants use a 12\:MB nursery for DaCapo
and Pjbb, and a 96\:MB nursery for the GraphChi applications. The reason to
use \kgb is to understand if simply using large nurseries, equal to the sum of nursery
and observer space in \kgw, could reduce PCM write rates similar to \kgw.

For the GraphChi applications, we evaluate \kgn and \kgb with the Large Object
Optimization (LOO) to form \kgn + LOO and \kgb + LOO.  We include the original
\kgw and two variants:  one that removes LOO to form \kgw--LOO and one that
removes the MetaData Optimization (MDO) to form \kgw--MDO.  We configure the
\wrgc collectors to have the observer space twice as large as the nursery.
Prior work reports this to be a good compromise between tenured garbage and
pause time. 

We compare to \pcm with the baseline generational Immix
collector~\cite{immix-blackburn}. We configure the baseline collector similar
to prior work~\cite{PLDI:2018:Shoaib}. All our experiments use two garbage
collector threads. 

\paragraph*{\textbf{Nursery and Heap Sizes}}
Nursery size has an impact on performance, response time, and space
efficiency~\cite{Appel,mmtk,Ungar:1992,Zhao:2009:AWL}. Similar to prior
work~\cite{PLDI:2018:Shoaib}, we use a nursery of 4\:MB for DaCapo and Pjbb.
Although recent prior work uses a 4\:MB nursery for GraphChi applications, we
find a 32\:MB nursery improves performance, and we use this size for our
experiments with GraphChi applications. We use a modest heap size that is twice
the minimum heap size. Our heap sizes reflects those used in recent
work~\cite{Akram:2017:DEP,immix-blackburn,Sartor:2014:Scrubbing,Zhao:2009:AWL,yak-fang}.

\paragraph*{\textbf{Hardware Platform}}
Figure~\ref{fig:platform} shows the NUMA platform we use to emulate hybrid
memory.  Each socket contains one Intel E5-2650L processor with 8
physical cores each with two hyperthreads, for 16
logical cores. The platform has 132 GB of main memory.  Physical memory is
evenly distributed between the two sockets. We use all the DRAM channels in
both sockets. The 20\:MB LLC on each processor is shared by all cores. The
maximum bandwidth to memory is 51.2\:GB/s; more than the maximum bandwidth
consumed by any of our workloads. The two sockets are connected by QPI links
that support up to 8\:GT/s.  We use Ubuntu 12.04.2 with 3.16.0
kernel.

\vspace*{-0.4em}
\section{Results}
\label{sec:results}

This section evaluates \wrgc collectors using emulation.  We first compare
emulation results to results of simulation and architecture-independent
analysis. Emulation results match prior results boosting our confidence in our
newly proposed methodology.  We use emulation to explore a richer space of
software configurations and workloads.  This enables us to discuss previously
unseen writes to memory due to interference patterns in multiprogrammed
workloads, simpler heap organizations for graph applications, and the
implications of production datasets. We conclude this section with reporting
raw write rates and PCM lifetimes in years.

\vspace*{-0.4em}
\subsection{Quantitative Comparison of Evaluation Methodologies}

Each evaluation methodology for hybrid memories has its strengths and weaknesses. 
Simulation models real hardware features but limits evaluation to a
few Java applications. Architecture-independent (ARCH-INDP) studies are fast
and improve application diversity but assume unrealistic hardware.  
ARCH-INDP counts the writes to virtual memory without taking
cache effects into account~\cite{PLDI:2018:Shoaib}.  Our goal in this section
is to explore if the major conclusions hold regardless of the evaluation
methodology, including the reduction in PCM write rates.  Lacking PCM hardware,
we can not compare accuracy. 

We first compare emulation results to simulation results. We reproduce
simulation results from previous work~\cite{PLDI:2018:Shoaib}. Lack of
full-system support and long simulation times limit evaluation using the
simulator to 7 DaCapo benchmarks: {\sf lusearch}, {\sf lu.Fix}, {\sf avrora},
{\sf xalan}, {\sf pmd}, {\sf pmd.S}, and {\sf bloat}. We use two configurations
of simulated hardware: (1) 4 cores and 4\:MB LLC, and (2) 4 cores and 20\:MB
LLC, which more closely matches the emulation platform.  

Table~\ref{fig:sim} shows the percentage reduction in PCM write rates reported
by the three methodologies.  Intuition suggests \wrgc collectors should be more
effective with a smaller LLC. Smaller LLC absorbs fewer writes which increases the
writes to PCM memory.  The results from two simulated systems in
Table~\ref{fig:sim} confirm this intuition.
 
LLC pressure is high on the simulated system with a 4\:MB LLC and the emulation
platform running a multiprogrammed workload with four instances. The nursery
size of the seven simulated benchmarks is 4\:MB, and the nursery is highly
mutated, so using an emulation system with four applications creates more LLC
interference and thus more PCM writes, similar to the simulated system with a
4\:MB LLC.  The average reduction in PCM write rate for both cases is similar.  

Simulated results with a 4\:MB LLC are different from emulation results with
one and two program workloads. This is because the nursery is a major source of
writes which are absorbed by the 20\:MB LLC in the emulation platform. 

We observe that simulation results for \kgn and a 20\:MB LLC report a 4\%
reduction in PCM writes. On the contrary, emulation results report a 29\%
reduction in PCM writes. This discrepancy is due to full-system effects in the
emulation platform. When emulating \kgn, we place explicit memory
allocations by the C and C++ libraries in DRAM. In \pcm, these
allocations are placed in PCM, which is why emulation reports a larger
reduction in PCM writes for \kgn. 

\begin{table}[bt]
 	\centering
	\includegraphics[width=8cm]{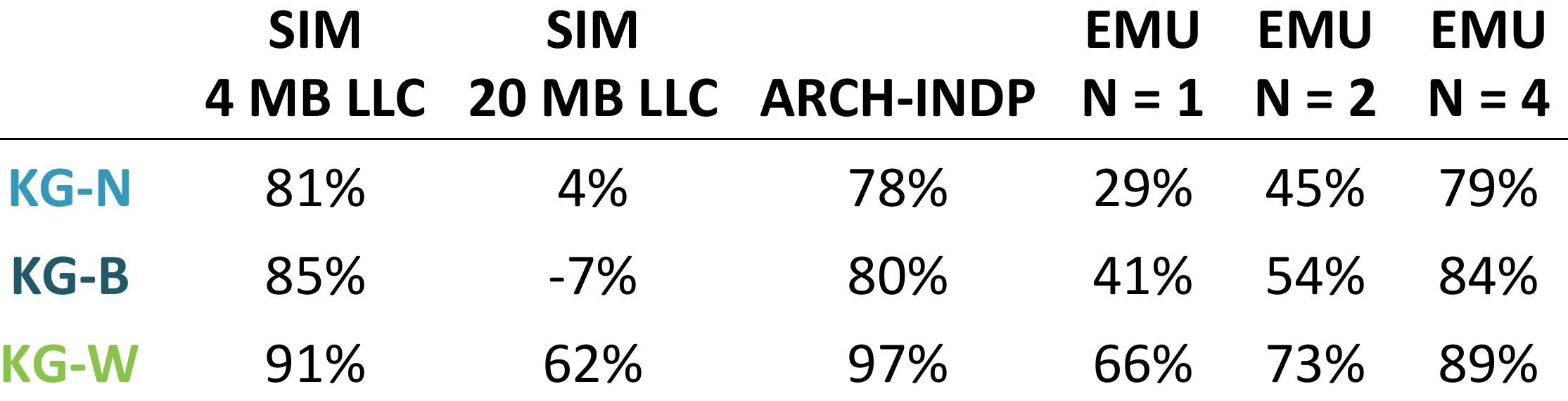}
	\caption{Comparing PCM write reduction using
simulation (SIM), architecture independent (ARCH-INDP) analysis, and emulation
(EMU) on 7 simulated benchmarks. N is the number of program instances in our multiprogrammed workloads. 
\textit{Simulation results confirm emulation results. The differences are due to
        cache sizes and full system effects.}
        }
	\label{fig:sim}
 	\centering
	\includegraphics[width=8cm]{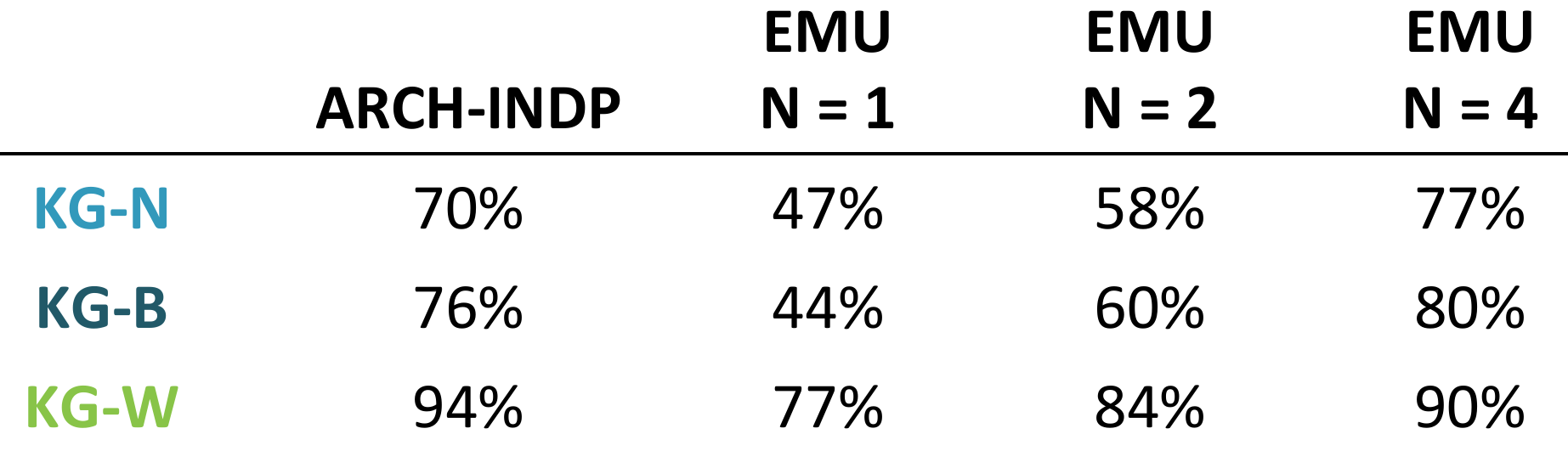}
	\caption{Comparing PCM write reduction using
ARCH-INDP and EMU for
all benchmarks. \textit{ARCH-INDP over-reports the reduction in
PCM writes for 1 and 2 program workloads.  For a  balanced workload that
utilizes all cores on Socket 0, EMU results match ARCH-INDP results.}
        }
	\vspace*{-2.2ex}
	\label{fig:archi}
\end{table}

We observe another discrepancy between the simulation and the emulation results for
\kgb with a 20\:MB LLC. Simulation results suggest increasing the nursery size
to 12\:MB leads to more PCM writes compared to a 4\:MB nursery.  On the
other hand, emulation results suggest a 41\% reduction in PCM write rate.  This
discrepancy opens up opportunities for future investigations. Bugs in one or
both environments could misreport PCM writes with larger nurseries.
Alternatively, the discrepancy could arise because of full system effects --
the simulation environment isolates Java heap allocations from OS and native
library allocations. We leave investigating this discrepancy further to future
work.

ARCH-INDP results over predict reductions when compared to emulation results
with a single program instance. They over predict because ARCH-INDP counts
successive writes to PCM virtual memory as writes to PCM physical memory. In
reality, some of those writes are filtered by the CPU caches. When workloads
exhibit large LLC interference, such as multiprogrammed workloads with four
instances, emulation results are in the ballpark of ARCH-INDP. We observe the
same behavior when comparing ARCH-INDP and emulation results for all benchmarks
in Table~\ref{fig:archi}.

\noindent\textbf{Finding 1.} \textit{LLC size impacts PCM write rates.
Simulation and emulation results converge with similar nursery to LLC size
ratios.  Full-system effects may cause discrepancies for some configurations.} 

\noindent\textbf{Finding 2.} \textit{Architecture-independent metrics
over-report the reduction in PCM write rates.}

\vspace*{-0.4em}
\subsection{Workload Analysis Using Emulation}

This section evaluates write-rationing garbage collectors much more fully than
prior work.  All trends previously observed for a narrow set of applications
and datasets are confirmed by emulation on a richer workload space. These
results make the case for using PCM as main memory even stronger. 

\paragraph{Garbage Collection Strategies for GraphChi}

Contrary to prior work, emulation provides us the opportunity to evaluate \wrgc
collectors for GraphChi applications. We find these applications allocate large
objects frequently.  We show that combining the optimization for large objects
in \kgw with the heap organization of \kgn is effective.  This configuration
simplifies heap management and improves performance. GraphChi applications have
more mature space collections than DaCapo and Pjbb. We tease apart the impact
of the metadata optimization here by evaluating \kgw--MDO.  We also
evaluate \kgw--LOO to tease apart the impact of the large object optimization
from \kgw. 

\begin{figure}[bt]
 	\centering
	\includegraphics[width=8cm]{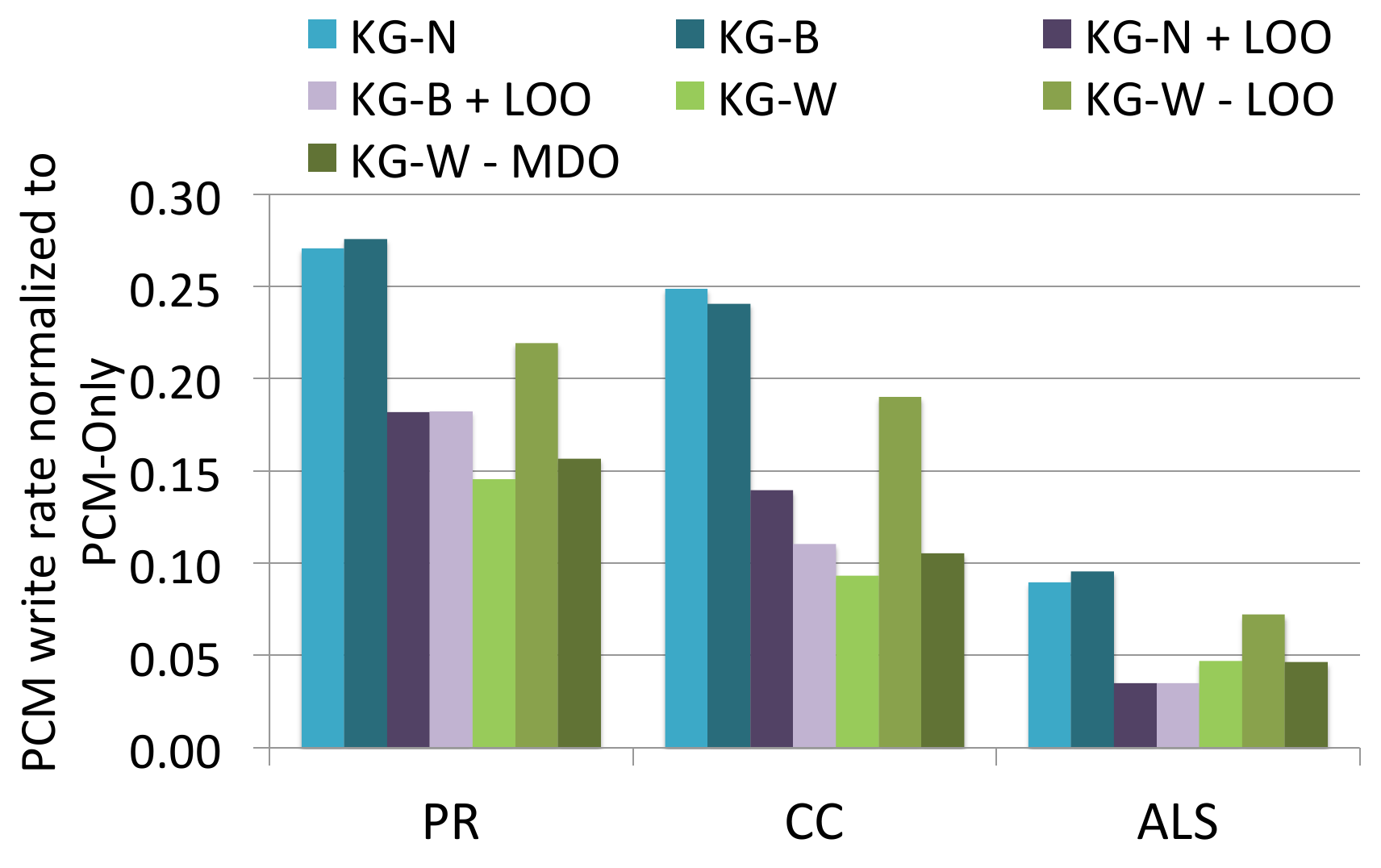}
	\caption{PCM write rates with various \wrgc collectors normalized to \pcm for GraphChi applications.  
		 \textit{\kgb + LOO works well for graph applications.}
        }
\vspace*{-1.7ex}
	\label{fig:graph-analysis}
\end{figure}

We show the write rates for single-program workloads normalized to \pcm in
Figure~\ref{fig:graph-analysis}. The absolute write rates increase for
multiprogrammed workloads but the normalized trends remain the same.  GraphChi
applications benefit from \kgn and \kgb that allocate new objects in a DRAM
nursery. This reduces write rates by 74\%, 75\%, and 91\% for {\sf PR}, {\sf
CC}, and {\sf ALS} respectively. \kgb uses a bigger nursery compared to \kgn
but still reduces write rates similar to \kgn. This confirms previous findings
with simulated benchmarks that we need novel heap organizations and other
optimizations to reduce PCM write rates further.

The graph applications allocate large objects and some follow the generational
hypothesis and thus benefit from the LOO optimization.  \kgn + LOO and \kgb +
LOO both reduce write rates on top of \kgn and \kgb respectively.  \kgn + LOO
reduces write rate by up to an additional 11\% compared to \kgn. \kgp + LOO is
even more effective for {\sf PR} and {\sf CC}: a 3\% additional reduction in
write rate. \kgn + LOO and \kgb + LOO are effective and have smaller execution
time overhead compared to \kgw.  Nevertheless, \kgw reduces write rates to PCM
more than \kgb + LOO for {\sf PR} and {\sf CC}.

Excluding LOO from \kgw (\kgw--LOO) increases the write rate because of large
object allocation in PCM of short-lived objects. Large object allocation in PCM further fills up the heap
quickly leading to more frequent mature collections. Mature collections are a source
of PCM writes because of the collector updates to the object mark
states.  The write rate increases by 3.3$\times$ for {\sf PR}, 2.6$\times$ for
{\sf CC}, and 1.5$\times$ for {\sf ALS}. 

Without the metadata optimization (MDO), PCM write rates increase, proving that
eliminating metadata writes during mature collections is effective. With one
instance of {\sf PR}, the write rate increases by 1.32$\times$ for {\sf PR} and
1.13$\times$for {\sf CC}. MDO benefits multiprogrammed workloads even more (not
shown).

\noindent\textbf{Finding 3.} \textit{Graph applications allocate many large
objects that benefit greatly from Kingsguard collectors that use the large
object optimization.}

\paragraph{Interference in Multiprogrammed Workloads}

\begin{figure}[t]
        \centering
	\includegraphics[width=8cm]{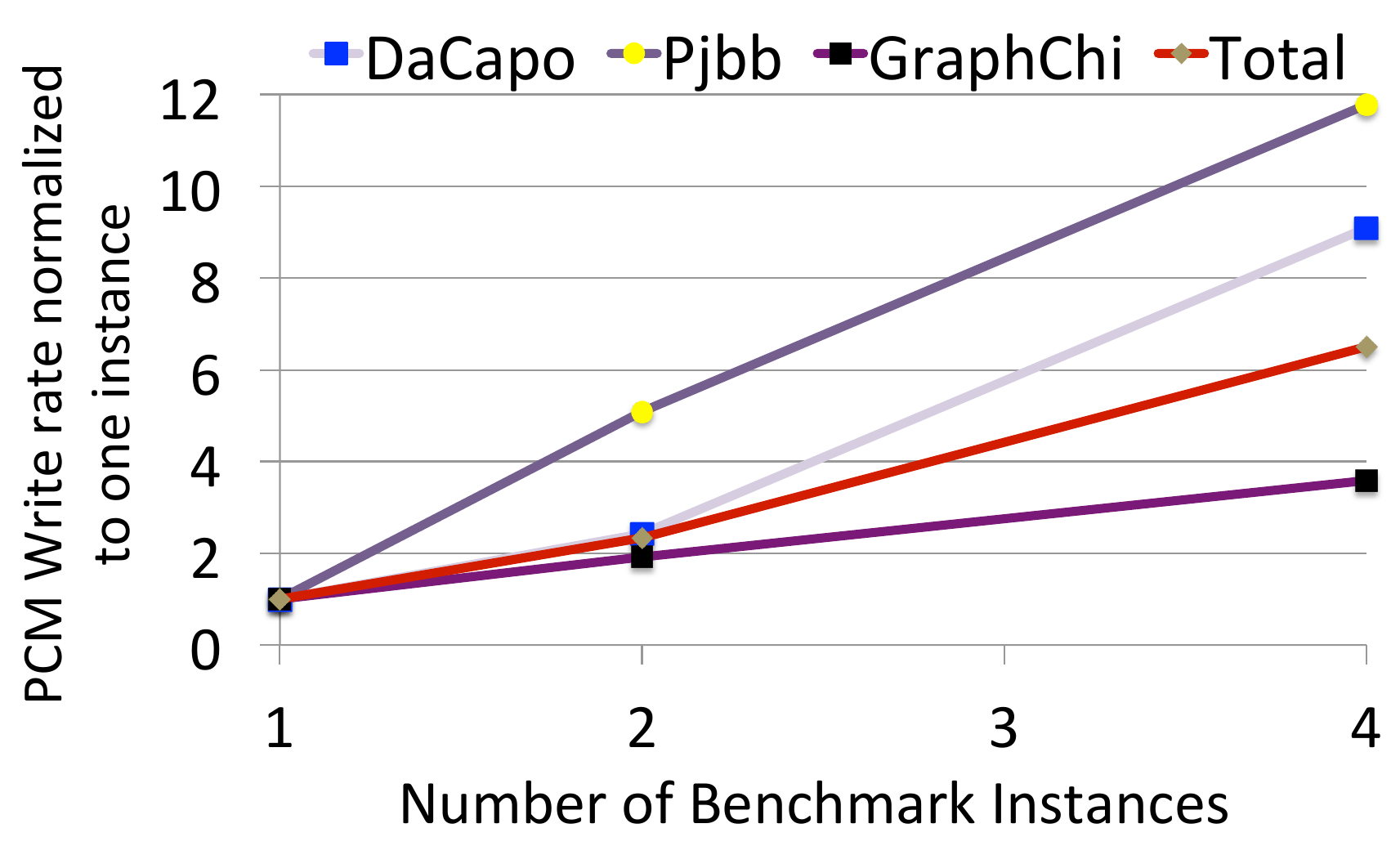}

	\caption{Average PCM write rates with \pcm 
normalized to single-program write rates.
\textit{Writes rates increase with the number of benchmark instances. The write rates of Pjbb and
DaCapo grow super-linearly from 1 to 4 instances.}}
	\label{fig:growthpcmonly}
        \vspace{2mm}
	\centering
	\includegraphics[width=8cm]{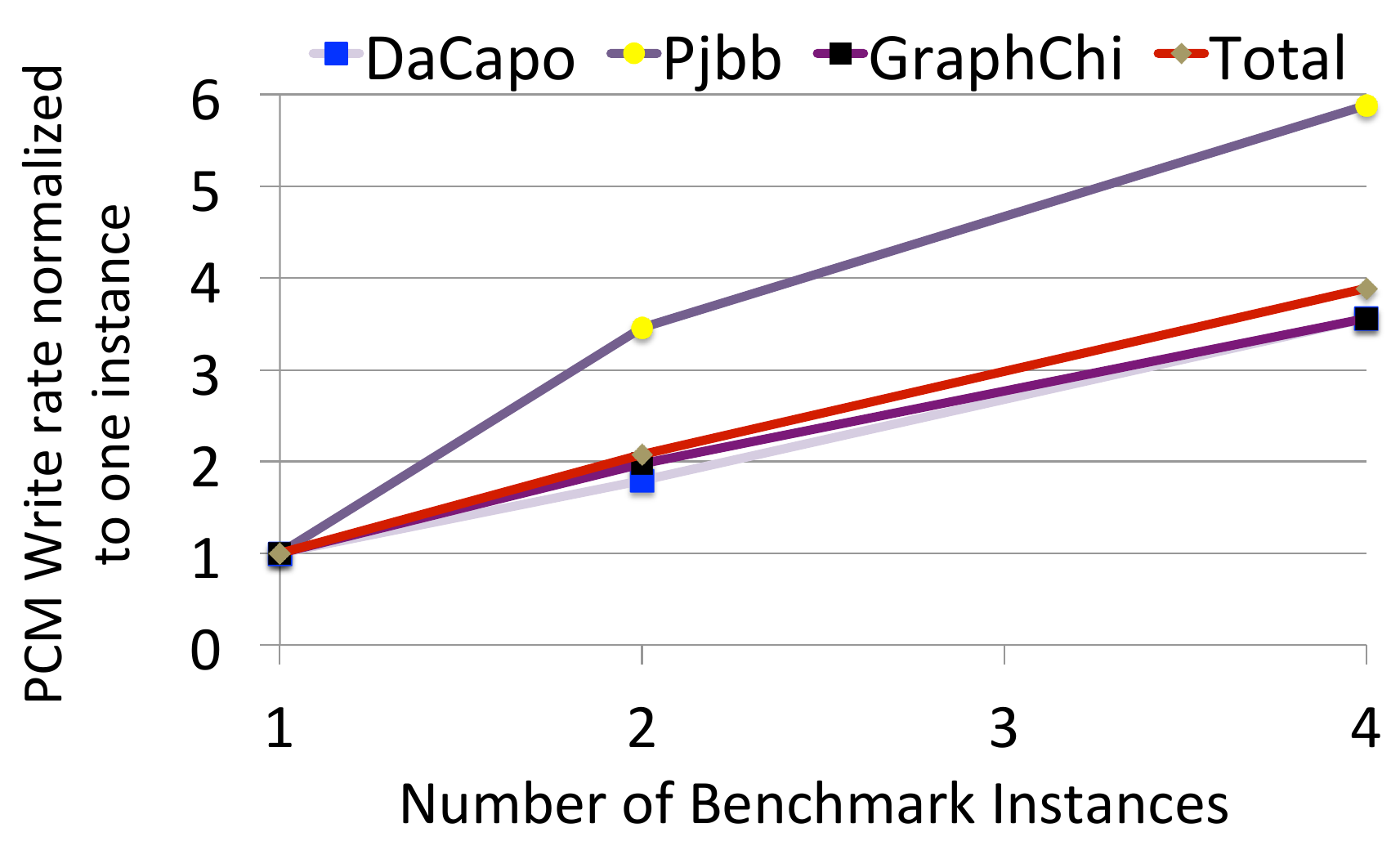}
	\caption{ Average PCM write rates with \kgw 
normalized to single-program write rates. \textit{The growth
in write rates is close to linear except Pjbb.} } 
	\label{fig:growthkgw}
\vspace*{-3.3ex}
\end{figure}

Long simulation times impede the evaluation of hybrid memories for
multiprogrammed Java workloads. The native execution speed of these workloads
on our emulation platform reveals interference patterns in the LLC which
results in writes to PCM memory.  Figure~\ref{fig:growthpcmonly} and
Figure~\ref{fig:growthkgw} shows the growth in average PCM write rates for \pcm
and \kgw for each benchmark suite and for all benchmarks.

We observe a variety of trends in write rates from the three suites.  On
average for \pcm, the increase in write rate from 1 to 2 program instances is
2.3$\times$, which is as we expect, but from 1 to 4 instances, the increase is
non-linear at 6.4$\times$.  DaCapo applications encounter high interference in
the LLC. The average increase in write rate from 1 to 4 instances for DaCapo is
9$\times$ (2.4$\times$ from 1 to 2 instances). The increase for Pjbb is even
higher. From 1 to 2 instances, the write rate increases by 5$\times$, and from
1 to 4 instances, the write rate increases by 12$\times$. GraphChi applications
on average show a linear trend. The increase in write rate from 1 to 4
instances is 1.9$\times$, and 3.5$\times$ from 1 to 4 instances.

Contrary to \pcm, \kgn and \kgw exhibit a linear increase in write rates from 1
to 2 and 4 program instances across the three suites: with 2 instances, the
increase is 1.8$\times$, 2.8$\times$, and 2.6$\times$ for DaCapo, Pjbb, and
GraphChi; with four instances, the increase is 3.1$\times$, 4.8$\times$, and
4.7$\times$ respectively. With \kgw, the increase is less than linear
except for Pjbb, which increases 6$\times$ with 4
program instances. 

\noindent\textbf{Finding 4.} \textit{PCM write rates grow super-linearly with
the number of concurrently running program instances for two popular Java
benchmark suites. Write rationing garbage collection significantly reduces the
growth in write rates.}

\paragraph{Modest versus Production Datasets}

\begin{figure}[bt]
 	\centering
	\includegraphics[width=8cm]{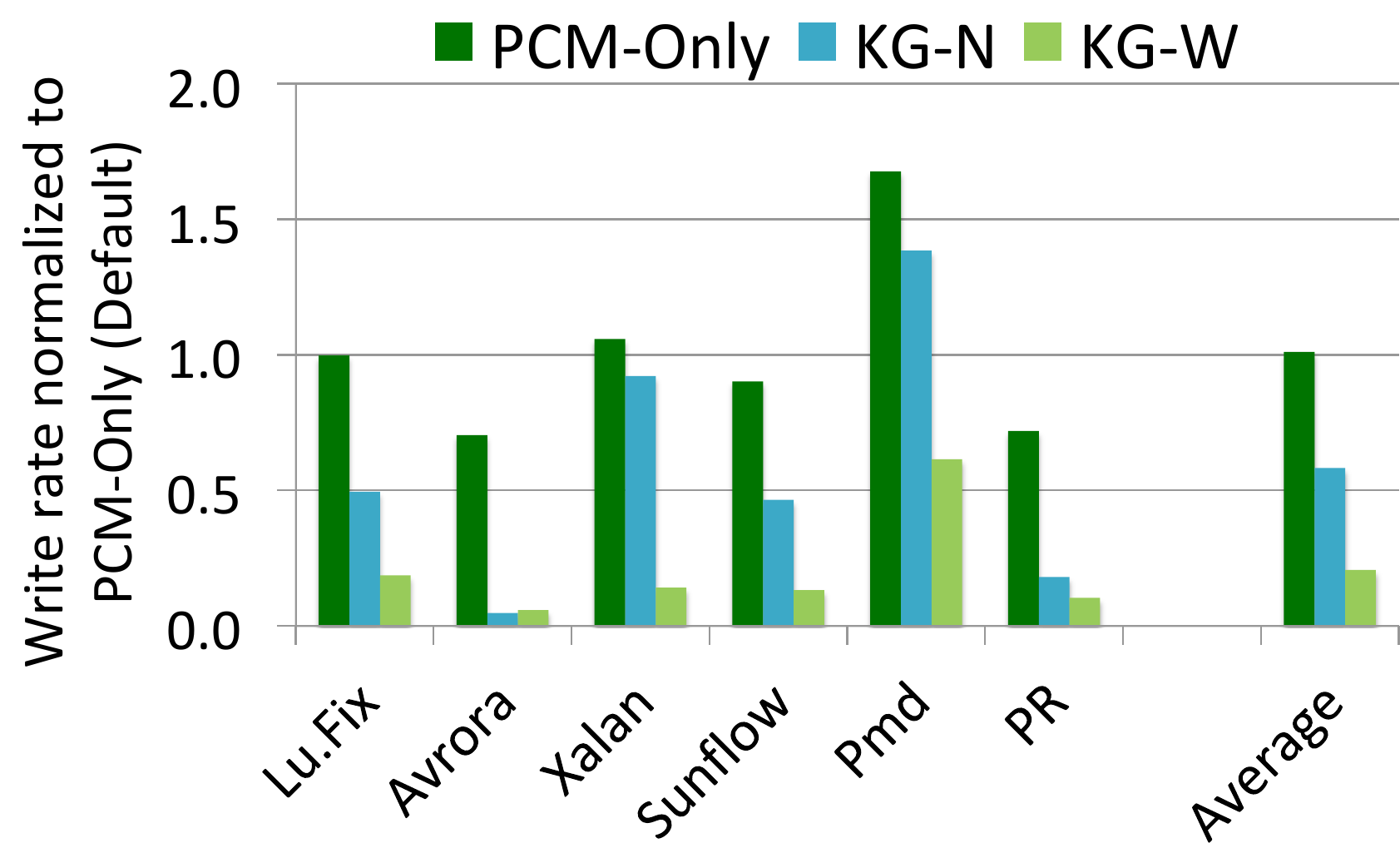}
	\caption{Write rates with large datasets normalized to write rates with
default datasets (\pcm). \textit{ Normalized write rates may change with
production datasets. PCM write rate still reduces with \kgn and
\kgw. }}
	\label{fig:dataset}
\vspace*{-2.2ex}
\end{figure}

The native speed of emulation admits larger datasets, previously unexamined.
The normalized writes rates for \pcm with large datasets shown in Figure~\ref{fig:dataset} 
follow three trends: The write rates of {\sf lusearch.fix} and {\sf xalan}
stay the same.  The write rates of {\sf avrora}, {\sf sunflow}, and {\sf PR} for
\pcm decrease by 0.7$\times$, 0.9$\times$, and 0.72$\times$ compared to default
datasets. The compute-to-write ratio of these applications increases with
larger inputs.  Conversely for {\sf pmd}, the write rate increases by
1.7$\times$.  Although absolute write rates change for some benchmarks, we
observe a similar reduction in PCM write rates with \kgn and \kgw. 

\noindent\textbf{Finding 5.} \textit{Production datasets sometimes shift the
balance between compute and memory-writes, changing PCM write rates.}
					
\paragraph{Classical versus Modern Suites}

\begin{figure}[bt]
 	\centering
	\includegraphics[width=8cm]{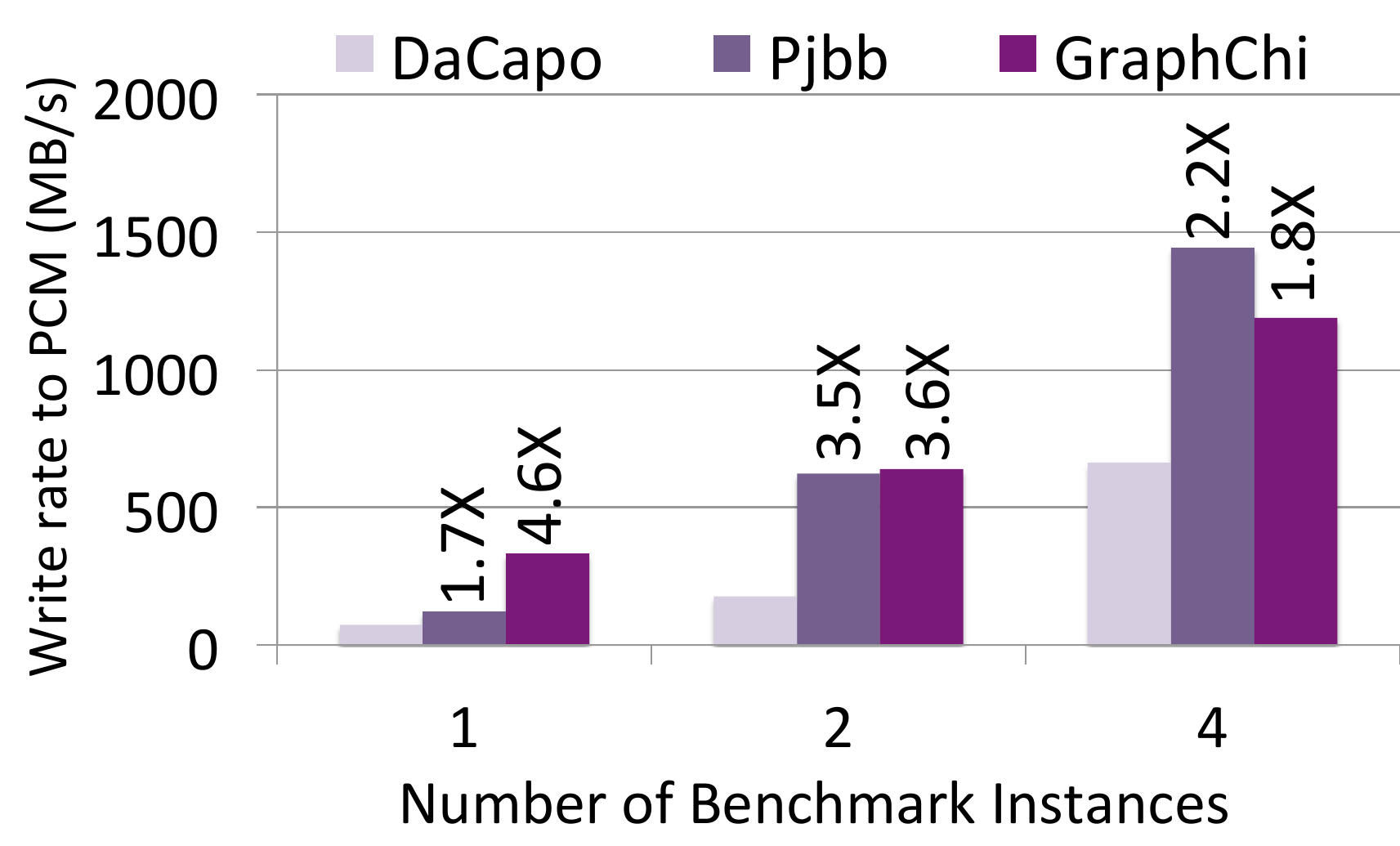}
	\caption{Average write rates in MB/s for DaCapo, Pjbb, and GraphChi
with \pcm. The numbers on top of Pjbb and GraphChi bars show the PCM write rate normalized to
DaCapo. \textit{Write rates increase for multiprogrammed
workloads. Pjbb and GraphChi have greater write rates compared to DaCapo.}}
	\label{fig:comp-suites}
\vspace*{-1.9ex}
\end{figure}

The DaCapo benchmark suite is the dominant choice for prior research in garbage
collection and some studies use Pjbb. We compare the average write rates of
Pjbb and GraphChi to DaCapo in Figure~\ref{fig:comp-suites}.
The average write rates of Pjbb and GraphChi are greater than DaCapo.  Pjbb
and GraphChi also have the largest heap sizes of all of our benchmarks.  The
average heap size of DaCapo is 100\:MB, Pjbb is 400\:MB, and GraphChi is
512\:MB. Pjbb has 1.7$\times$ the write rate of DaCapo with single-program
workloads.  Although we expect both Pjbb and GraphChi to have higher
write rates than DaCapo, it is interesting that GraphChi has a 4.7$\times$
higher write rate than DaCapo. 

\noindent\textbf{Finding 6.} \textit{Future studies on hybrid memories should
use a diversity of applications, including large heaps.}

The write rates of Pjbb and GraphChi are higher even for multiprogrammed
workloads. The difference is less pronounced compared to single-program
workloads because with four instances, DaCapo applications on average
experience greater interference in the LLC. The DaCapo rates increase
super-linearly whereas the increase is less than super-linear for Pjbb and
GraphChi. 

The difference in write rates between DaCapo and GraphChi is less pronounced
with \kgn and \kgw.  GraphChi has the same average write rate as DaCapo with
\kgn and single-program workloads. The write rate with two and four instances
is 1.5$\times$ and 1.6$\times$ higher than DaCapo. Thus, a large reason for
the gap in write rates of DaCapo and GraphChi for \pcm is the nursery writes.

The write rates of Pjbb with \kgn and \kgw are still higher compared to the
average DaCapo rates. For instance, with \kgw and single-program workloads, the
write rate of Pjbb is 3$\times$ that of DaCapo (2$\times$ for \kgn). For two
and four instances with \kgw, Pjbb incurs a 5.7$\times$ and 5$\times$ higher
write rate compared to DaCapo. 

\noindent\textbf{Finding 7.} \textit{Pjbb and GraphChi have higher write rates
than DaCapo.}

\paragraph{PCM Write Rates and Implications for Lifetime}
\label{sec:write-rates}

\begin{figure}[bt]
 	\centering
	\includegraphics[width=8cm]{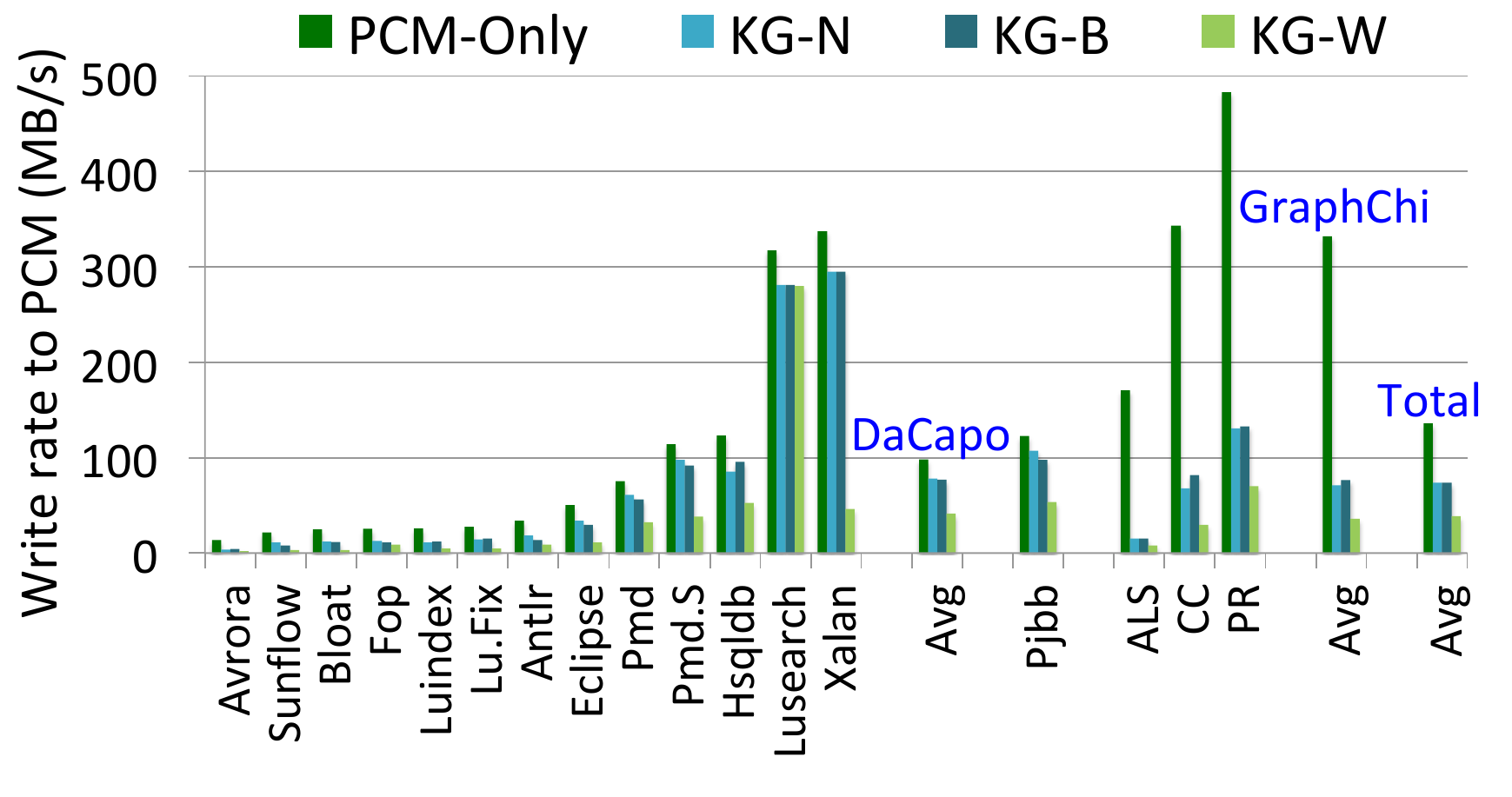}
	\caption{Write rates in MB/s for one instance workloads. 
		 \textit{Benchmarks exhibit a range of write rates. 
                         Applications that allocate large objects frequently have the highest write rates.}
        }
	\label{fig:abs-wr-1vm}
	\vspace{2mm}
 	\centering
	\includegraphics[width=8cm]{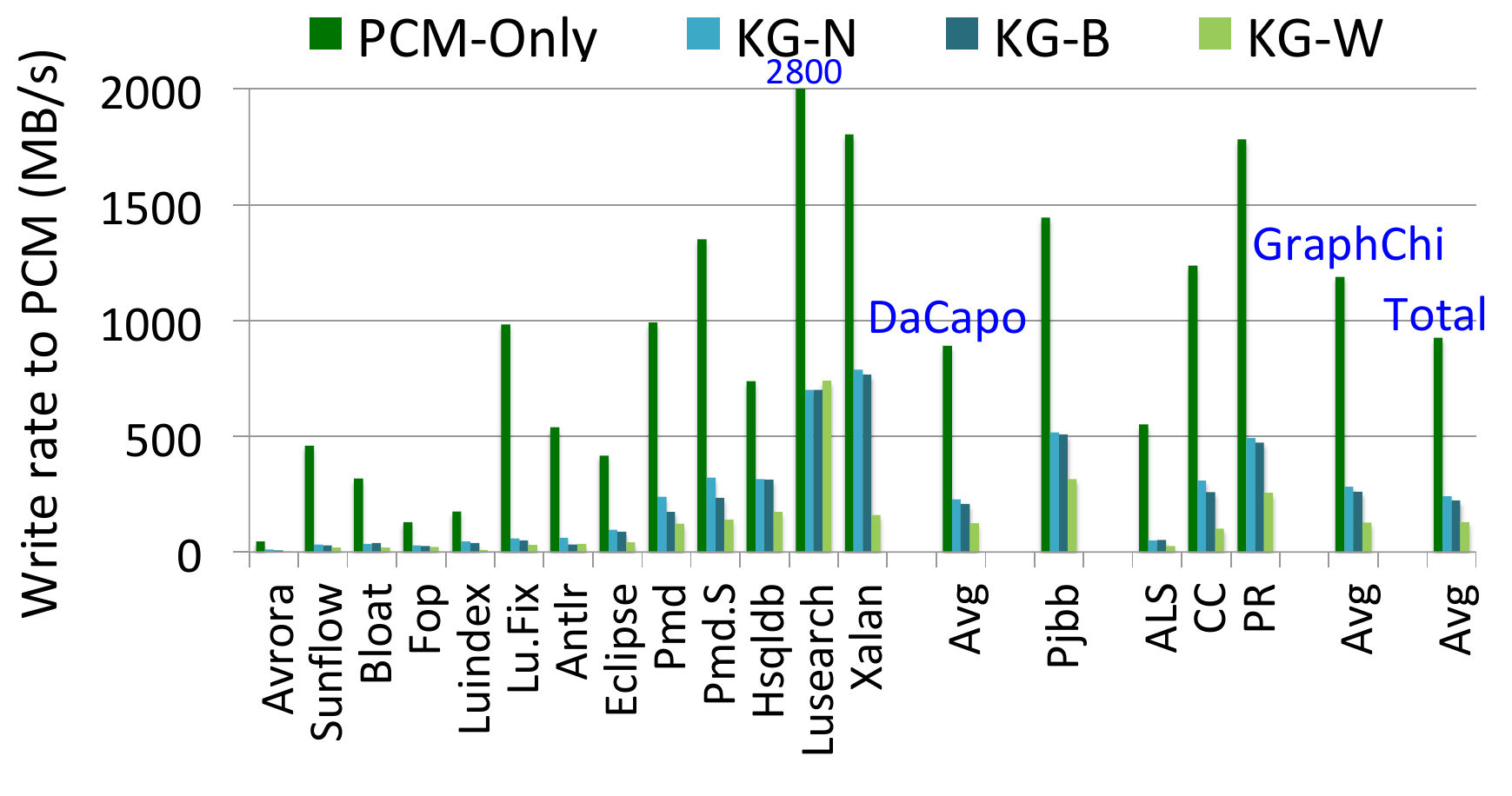}
	\caption{Write rates in MB/s with 4 instance workloads.
        \textit{Write rates reduce significantly across all benchmarks with \wrgc collectors.}
        }
	\label{fig:abs-wr-4vm}
\vspace*{-3.3ex}
\end{figure}

We now show the raw write rates to PCM for our benchmarks and reduction in
write rates using \kgw.  We also discuss PCM lifetime in years for our
workloads. Lifetime is a linear function of write rate and PCM cell endurance.
We compute PCM lifetimes similar to prior work assuming a PCM write endurance of
10\:M writes per
cell~\cite{lee-pcm-isca,moin-pcm-isca,Moin-Start-Gap,PLDI:2018:Shoaib}.  We
assume a 32\:GB PCM system with hardware wear-leveling that delivers endurance
within 50\% of the theoretical maximum~\cite{Moin-Start-Gap}.
Table~\ref{tab:lifetime} shows worst-case PCM lifetimes in years for the three
benchmark suites. We choose the shortest lifetime of all benchmarks for DaCapo
and GraphChi. We only consider the fixed version of {\sf lusearch} in the
worst-case lifetime analysis.

Figure~\ref{fig:abs-wr-1vm} shows the write rates for \pcm and three \wrgc
configurations for single programs.  The average PCM write rate for \pcm is
126\:MB/s and write rates vary from 14\:MB/s for {\sf avrora} to 480\:MB/s for
{\sf PR}. {\sf lusearch} is excluded from the average.  Higher write rates of
GraphChi applications limit memory lifetime of \pcm to only 10.5 years with
single programs. The worst-case lifetime in DaCapo is 14 years for {\sf xalan}.
Wear-leveling and write filtering by LLC alone can make PCM last for 41 years
when running a single instance of Pjbb.

\begin{table}[b]
        \centering
        \vspace*{-1.3em}
        \includegraphics[width=8cm]{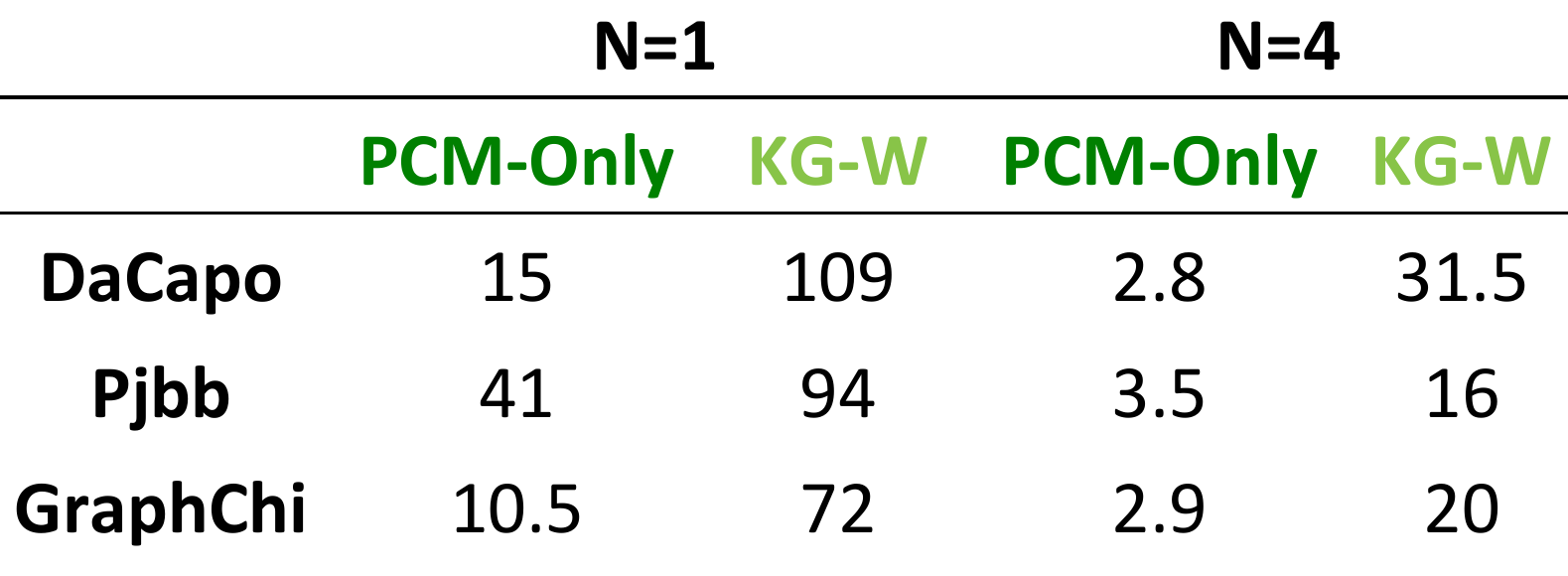}
        \vspace{0.5mm}
        \caption{PCM lifetime in years for single-program (N=1) and 4-program (N=4) workloads. 
\textit{GraphChi and multiprogrammed workloads quickly wear out PCM. \wrgc collectors make PCM practical for all
workloads.}
        }
        \label{tab:lifetime}
\end{table}

\noindent\textbf{Finding 8.} \textit{Graph processing applications wear out PCM
much more quickly than DaCapo and Pjbb.} 

Of all the DaCapo benchmarks, {\sf lusearch} has the highest write rate of
320\:MB/s. Interestingly, {\sf lusearch.fix} fixes an allocation bug in the
original {\sf lusearch} and has a write rate of only 27\:MB/s. We also observe
a change in write rate between the two versions of {\sf pmd}. The original
benchmark has a write rate of 75\:MB/s on our platform. The version of the
benchmark that removes an input file for better scaling with the number of
threads, {\sf pmd.S}, has a write rate of 114\:MB/s. The execution time of {\sf
pmd.S} reduces significantly compared to {\sf pmd} leading to this higher write
rate.  

A widely used application in the DaCapo suite, {\sf eclipse}, has a write rate
of 50\:MB/s. This write rate is less than transaction ({\sf hsqldb} and 
Pjbb) and graph processing applications.  On the other hand, it is higher than
applications that do lexical analysis such as {\sf antlr} and {\sf bloat}.
{\sf ALS} with 170\:MB/s has the lowest write rate of the three GraphChi
applications.

\noindent\textbf{Finding 9.} \textit{Applications from different benchmark
suites and from different domains within a suite exhibit a variety of PCM write
rates. Applications that allocate large objects abundantly have higher write
rates.} 

\noindent\textbf{Finding 10.} \textit{Allocation behavior and input sets
influence write rates.}

\wrgc collectors significantly reduce PCM write rates across the three
benchmark suites. \kgn reduces the
average write rate by 50\% for single programs. The average write rate of \kgn
is 60\:MB/s. \kgb with its bigger nursery results in the same PCM write rates as \kgn.
GraphChi applications write much less to PCM with \kgn.  This
shows that the nursery is highly mutated even in modern graph processing
applications.  The benchmarks that do not benefit a lot from \kgn are those
that: (1) profusely allocate large short-lived data structures such as {\sf
lusearch} and {\sf xalan}, and (2) have more mature-object writes than nursery
writes, such as Pjbb~\cite{PLDI:2018:Shoaib}. 

\noindent\textbf{Finding 11.} \textit{Simply using DRAM for larger nurseries
does not reduce PCM write rates in hybrid memory systems.}

\kgw reduces the average write rate by 80\% and the raw average write rate is 24\:MB/s.
These low write rates greatly improve PCM lifetime.  PCM lifetime with \kgw for
single programs is practical across all three benchmark suites. For instance,
GraphChi applications with \kgw will wear out PCM after 72 years. PCM will also
be used for persistent storage which can have many more writes. The lifetimes
shown in Table~\ref{tab:lifetime} are therefore optimistic.

Figure~\ref{fig:abs-wr-4vm} shows PCM write rates of 4-program workloads.
Write rates increase a lot and up to 2.8\:GB/s for {\sf lusearch}.
\kgn reduces the write rate significantly for the 4-program {\sf lusearch}
workload.  The increase in write rates for multiprogrammed workloads has
implications for PCM lifetimes.  The average lifetime for the DaCapo suite
is not even 5 years. Lifetimes of Pjbb and GraphChi are worse.

\noindent\textbf{Finding 12.} \textit{Multiprogramming workloads can
wear out PCM memory in less than 5 years.}

\begin{figure}[bt]
 	\centering
	\includegraphics[width=8cm]{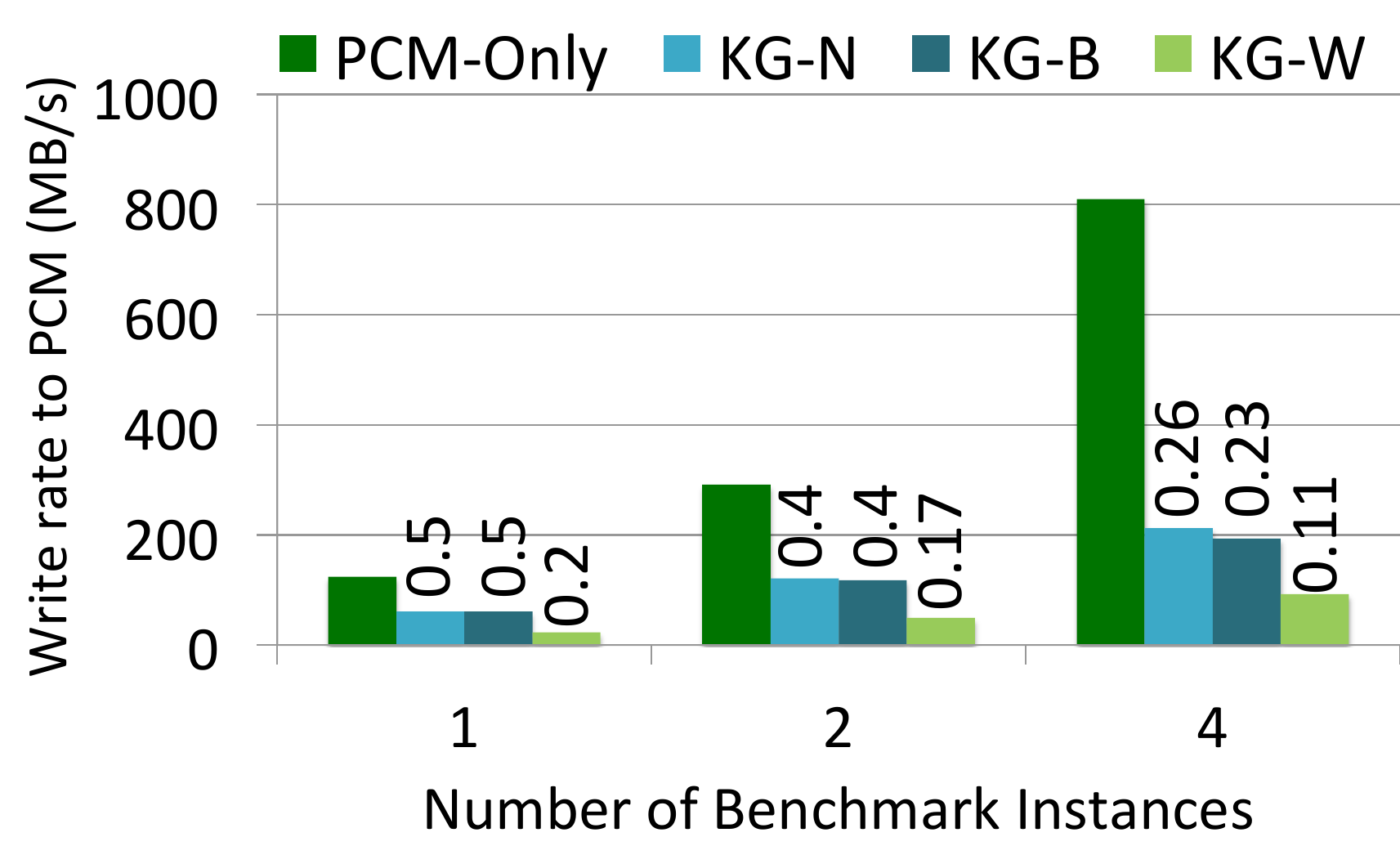}
	\caption{Average write rates in MB/s with varying number of benchmark
instances. The numbers on top of \kgn, \kgb, and \kgw bars show write rates
normalized to \pcm. \textit{Write rates increase with increasing number of
benchmark instances. \wrgc collectors are more effective for multiprogrammed
workloads.}}
	\label{fig:abs-wr}
\vspace*{-3.0ex}
\end{figure}

Figure~\ref{fig:abs-wr} compares average write rates of single-program and
multiprogrammed workloads with \pcm, \kgn, and \kgw.  Write rates reach close
to 1\:GB/s with two program instances and up to 2.8\:GB/s with four program
instances. Fortunately, write rates to PCM drop to less than 100\:MB/s on
average with \kgw across all workloads.  Figure~\ref{fig:abs-wr} also shows
that \kgn, \kgb, and \kgw are more effective in normalized terms for
multiprogrammed workloads.  Write rates increase due to interference in the LLC
and most of the interference is due to nursery writes.  Contrary to \kgn
reducing the write rate to PCM by 50\% with one instance, \kgn reduces the
write rate to PCM by 80\% with 4 program instances.  

\noindent\textbf{Finding 13.} \textit{Concurrently running applications in
multiprogrammed environments incur LLC interference due to nursery writes.
\wrgc collectors are highly effective in such environments.}

Table~\ref{tab:lifetime} shows that \wrgc collectors bring PCM lifetimes to
practical levels for multiprogrammed workloads. The worst-case lifetime is more
than 15 years for DaCapo, Pjbb, and GraphChi. Software and hardware approaches
together can make PCM a practical replacement for DRAM.

\paragraph{Execution Time}

Overall across single-program and multiprogrammed workloads and compared to
\kgn, \kgb slightly reduces the execution time, and \kgw increases the
execution time. The average reduction with \kgb is 3\%, and average increase
with \kgw is 10\% for single programs. The results are in the ballpark for
multiprogrammed workloads.  {\sf hsqldb} suffers the highest overhead of 28\%.
An exception with \kgw is {\sf bloat} whose execution time reduces up to 12\%.
The low survival rate of observer collections leads to fewer mature collections
which improve overall application performance.

\noindent\textbf{Finding 14.} \textit{There is a price to pay for severely
limiting writes to PCM.  KG-W's overhead ranges from 0-28\%.}

\vspace*{-0.4em}
\section{Related Work}
\label{sec:related}

Now we discuss related work on methods to evaluate hybrid memories, and
managed runtimes for emerging hardware.

\vspace*{-0.6em}
\subsection{Evaluation methodologies}
\vspace*{-0.2em}

Prior work uses emulation to evaluate emerging
memories~\cite{Oskin:2015:SAD,Wang:2016,Dulloor:2014:SSP}. Oskin et al.  use a
NUMA platform for emulating die-stacked DRAM~\cite{Oskin:2015:SAD}. Their
evaluation only considers applications written in C. Dulloor et al. emulate
hybrid DRAM-NVM memory on a NUMA platform but use it to evaluate filesystems
for persistent object storage.   

Two platforms today enable executing Java applications on top of simulated
hardware in a reasonable time: (1) Jikes RVM on top of
Sniper~\cite{Sartor:2014:Scrubbing}, and (2) Maxine VM on top of
ZSim~\cite{maxine:ispass}. Both Sniper and ZSim are a cycle-level multicore
simulators that trade off some accuracy for speedy
evaluations~\cite{carlson2014aeohmcm,Sanchez:2013:ZFA}. In their publicly
available versions, both platforms lack support for full-system simulation;
favoring speed over detail. 

Cao et al. use emulation to evaluate managed runtimes for hybrid memories but
their infrastructure only supports simple heap organizations. Our platform is
flexible and enables the evaluation of a range of collector configurations. We
also provide a methodology to measure write rates of Java workloads that are
run using replay compilation~\cite{ha_et_al_2008,huang_et_al_2004}.

\vspace*{-0.5em}
\subsection{Managed runtimes for emerging hardware}

Prior work has looked into tailoring the managed runtime for
hybrid memories. Wang et al. use DRAM in hybrid DRAM-NVM systems for allocating
frequently read objects~\cite{Wang:2016}. They use an offline profiling phase
to identify hot methods in the program. During runtime, all object allocation
that happens from hot methods goes into DRAM.  Unlike write-rationing
collectors that target lifetime, their goal is performance. 

Gao et al., use the managed runtimes to tolerate PCM
failures~\cite{Gao:2013:UMR}. The hardware informs the OS of defective lines
which in turn communicates faulty lines to the garbage collector. The garbage
collector masks the defective lines and moves data away from them.

We discussed write-rationing garbage collection for hybrid
memories~\cite{PLDI:2018:Shoaib} proposed by Akram et al., in
Section~\ref{sec:background}.  Recent work predicts write-intensive objects
using offline profiling to reduce the overheads of online monitoring
~\cite{PASS:2018:Shoaib}.

\ignore{
Apart from memories, prior work uses scheduling to optimize the performance and
energy of managed applications on heterogeneous
multicores\cite{taco-akram-2016,yinyang,Jibaja:2016:PPA}. \ignore{These works
investigate how best to schedule application and runtime services on small
versus big cores.} 
}

\vspace*{-1.3ex}
\section{Conclusions}
\label{sec:concl}

\vspace*{-0.4ex}
Advances in non-volatile memory (NVM) technologies have implications for the
whole computing stack. Researchers need fast and accurate methodologies for
evaluating NVM as memory and storage. This work introduces an emulation
platform built using widely available NUMA servers to accurately measure read
and write rates and the performance of hybrid memories that combine DRAM and
NVM. This platform can be used to evaluate applications that use manual or
automatic memory management. We evaluate our emulator with write-rationing
garbage collectors that keep frequently written objects in DRAM to guard NVM
against writes. We compare emulation to simulation and architecture-independent
analysis, showing they have similar trends. With the emulation, we can
and do explore large graph applications and multi-programmed workloads
with large datasets.  Emulation reveals new insights, such as that modern graph
applications have much larger write rates than DaCapo, and benefit greatly from
write-rationing collectors.  Multiprogrammed environments see a super-linear
growth in write rates compared to running single programs. This growth goes
away with write-rationing collectors.  Although simulation and emulation both
have their place, emulation adds the ability to explore a richer
software design and workload space. 


\bibliographystyle{IEEETran}
\bibliography{paper}

\end{document}